\documentclass[a4paper,fleqn]{cas-dc}

\usepackage[numbers,sort&compress]{natbib}

\usepackage{amssymb}
\usepackage{amsmath}
\usepackage{siunitx}
\usepackage{gensymb}
\usepackage{hyperref}
\usepackage[T1]{fontenc}

\usepackage{xcolor}
\usepackage{ulem}
\usepackage{soul}

\usepackage{algorithm}
\usepackage[noend]{algpseudocode}

\newcommand{\rev}[1]{#1}
\begin{document}
\let\WriteBookmarks\relax
\def\floatpagepagefraction{1}
\def\textpagefraction{.001}

\shorttitle{Stochastic simulation of partial discharge inception}

\shortauthors{J. Teunissen \& Y. Gao}

\title [mode = title]{Stochastic simulation of partial discharge inception}



%

\author[1,2]{Jannis Teunissen}[
orcid=0000-0003-0811-5091
]



\ead{jannis.teunissen@cwi.nl}


\credit{Conceptualization of this study, Funding acquisition, Methodology, Software, Visualization, Writing – original draft, Supervision}

\affiliation[1]{organization={Centrum Wiskunde \& Informatica (CWI)},
    addressline={Science Park 123},
    city={Amsterdam},
    postcode={1098 XG},
    country={The Netherlands}}

\affiliation[2]{organization={Centre for mathematical Plasma Astrophysics, Department of Mathematics, KU Leuven},
    addressline={Celestijnenlaan 200B},
    city={Leuven},
    postcode={3001},
    country={Belgium}}

\author[1]{Yuting Gao}[
orcid=0009-0008-3644-9658
]


\ead{yuting.gao@cwi.nl}


\credit{Conceptualization of this study, Investigation, Methodology, Software, Visualization, Writing – review and editing}

\cortext[1]{Corresponding author}



\begin{abstract}
    We present a Monte Carlo method for simulating the inception of electric discharges in gases.
    The input consists of an unstructured grid containing the electrostatic field.
    The output of the model is the estimated probability of discharge inception per initial electron position, as well as the estimated time lag between the appearance of the initial electron and discharge inception.
    To obtain these quantities electron avalanches are simulated for initial electron positions throughout the whole domain, also including regions below the critical electric field.
    Avalanches are assumed to propagate along field lines, and they can produce additional avalanches due to photon and ion feedback.
    If the number of avalanches keeps increasing over time we assume that an electric discharge will eventually form.
    A statistical distribution for the electron avalanche size is used, which is also valid for gases with strong electron attachment.
    We compare this distribution against the results of particle simulations.
    Furthermore, we demonstrate examples of inception simulations in 2D Cartesian, 2D axisymmetric and 3D electrode geometries.
\end{abstract}



\begin{keywords}
  partial discharge \sep electric discharge \sep Monte Carlo simulation \sep inception voltage
\end{keywords}

\maketitle

\section{Introduction}
\label{sec:introduction}

The term ``partial discharges'' \rev{(PD)} is typically used for electric discharges confined by an insulating medium~\cite{VanBrunt_1991,Niemeyer_1995}.
Understanding under what conditions such discharges form is important, since they can degrade the insulating material and lead to \rev{equipment} failures.
In general, a region of high electric field is required in which the electron impact ionization coefficient $\alpha$ exceeds the electron attachment coefficient $\eta$, so that electron avalanches can form.
An initial electron is also required, and to sustain the avalanches a secondary electron emission (SEE) mechanism is necessary, such as photoionization, or secondary emission due to the impact of ions or photons on surfaces.

An overview of experimental techniques to detect partial discharges can be found in e.g.~\cite{Bartnikas_2002,Pattanadech_2023}, but here we focus on modeling their inception.
A traditional approach has been to consider the ionization integral along a field line
\begin{equation}
  \label{eq:K-integral}
  K = \int \alpha_{\mathrm{eff}}(x') \, dx',
\end{equation}
where $\alpha_{\mathrm{eff}} = \alpha - \eta$ is the effective ionization coefficient.
A measured partial discharge inception voltage (PDIV) can then be associated with a threshold value for $K$.
A wide range of $K$ values has been obtained in experiments, mostly caused by differences in electric field non-uniformity and in SEE processes~\cite{Farber_2023}.
When the SEE processes and their coefficients are known, a $K$ threshold can computationally be obtained in a given geometry.
For example, in~\cite{Naidis_2005} criteria for positive discharge growth in air around spherical and cylindrical electrodes were obtained, with photoionization as the SEE mechanism.

A threshold for $K$ can be related to an effective SEE coefficient $\gamma_\mathrm{eff}$ as~\cite{Farber_2023}
\begin{equation}
  \label{eq:gamma-eff}
  \gamma_\mathrm{eff} = \exp(-K),
\end{equation}
where $\gamma_\mathrm{eff}$ is the average number of secondary electrons emitted per ionization in the primary avalanche.
In general, $\gamma_\mathrm{eff}$ can depend on several factors, such as the gas used, the surface properties of materials, and the considered geometry.
Furthermore, it should be noted that for gases with strong electron attachment the number of ionization events can significantly exceed $\exp(K)$.
For the homogeneous electric fields considered in~\cite{Farber_2023} this was corrected for by multiplying the right-hand side of equation~\eqref{eq:gamma-eff} by a factor $\alpha_\mathrm{eff}/\alpha$.
Here we will instead define a custom ionization integral $K^*$ that takes attachment into account, as discussed in section~\ref{sec:numb-ioniz-avalanche}.

So-called fluid or particle models can be used to simulate the evolution of electrons (and other relevant species) in time, including their effect on the electric field.
In the context of partial discharges, such models have been used in e.g.~\cite{Mikropoulos_2016,Callender_2018,Marskar_2025}.
SEE mechanisms can directly be incorporated in these models, and they can be used to study the temporal and spatial evolution of a discharge, for example the avalanche-to-streamer transition~\cite{Nijdam_2020}.
However, conventional electric discharge models are typically too expensive to perform runs with many initial electron positions and applied voltages.

With a fluid model it is also possible to search for steady-state solutions~\cite{Benilov_2021}.
The PDIV can then be found by as the steady-state solution at which the production and loss of electrons is balanced.
Since the charged particle densities at the inception voltage are low, a static (Laplacian) electric field can be used, greatly simplifying the problem.
The resonance method described in~\cite{Almeida_2020,Benilov_2021} seems promising in this regard.

For the design of equipment in which PDs are to be prevented, simple and computationally efficient methods to predict the PDIV are desired.
In~\cite{Marskar_2025} an approach was presented in which $K$ is compared against two thresholds: the first being $K_c$, a threshold for streamer formation, and the other indicating sustained discharge growth through ion-induced SEE.
The predictions of this approach were found to be consistent with `It\^o-KMC' simulations.
In~\cite{Korthauer_2025} a similar approach was presented, but now using a parametrization of the effective SEE coefficient $\gamma_\mathrm{eff}$ based on~\cite{Farber_2023} and including a correction to $K_c$ depending on the size of the region where $\alpha_{\mathrm{eff}} > 0$.
Good agreement was found between the PDIV predicted by this model and experimental measurements in different geometries.
Several other simplified models have also been used to predict the onset of partial discharges, see for example~\cite{Niemeyer_1995,Pan_2019}.

All the approaches described above are deterministic: they indicate whether a discharge can form or not, but they do not provide information on the probability or time delay of inception.
Here we present a stochastic model for avalanche growth, in which we do not simulate individual electrons but whole avalanches at a time.
This makes it possible to efficiently estimate the probability of discharge inception for different initial electron positions at a given applied voltage.
Furthermore, the delay between the appearance of a free electron and the formation of a discharge can be estimated.
In the present paper, we describe the stochastic model and its implementation, and we show several examples.
A detailed comparison against experimental measurements is left for future work.

\section{Model}
\label{sec:model-description}

The components of the model are described below, and its implementation is discussed in section~\ref{sec:implementation}.
It is important to note that we use `avalanche size' for the total amount of ionization $M$ produced by an avalanche.
The number of electrons in an avalanche at some position or time is indicated by the symbol $N$, as discussed in sections~\ref{sec:electr-aval-num} and \ref{sec:numb-ioniz-avalanche}.

\rev{Most processes in the model are handled stochastically, such as the sampling of avalanche sizes and the generation of secondary electrons.
  However, the propagation of avalanches and ions is assumed to occur deterministically along electric field lines, and diffusive effects due to collisions are not taken into account.}

\subsection{Distribution for the number of electrons in an avalanche}
\label{sec:electr-aval-num}

Different distributions and approximations for the number of electrons in an avalanche have been proposed,
as reviewed in~\cite{VanBrunt_1991}.
Here we simply assume that $\alpha$ and $\eta$ are both functions of the local electric field strength $E$, which we for brevity express as $\alpha(x)$ instead of $\alpha[E(x)]$.
In 1948 Kendall derived general probability distributions for birth and death processes~\cite{Kendall_1948}.
These results can be applied to avalanche growth by replacing the birth rate with $\alpha$, the death rate with $\eta$, and the temporal dependence with a spatial one.

The main findings from~\cite{Kendall_1948} relevant for this paper are summarized below.
Starting from a single electron at $x_0$, the expected number of electrons in an avalanche at a location $x$ is
\begin{equation}
  \label{eq:mean-func}
  \bar{N}(x) = e^{K(x)},
\end{equation}
with $K(x)$ defined as in equation~\eqref{eq:K-integral}
\begin{equation}
  \label{eq:K-func}
  K(x) = \int_{x_0}^{x} [\alpha(x') - \eta(x')] \, dx'.
\end{equation}
The probability that the avalanche contains zero electrons is
\begin{equation}
  \label{eq:p0}
  P_0(x) = 1 - \bar{N}(x)/W(x),
\end{equation}
where
\begin{align}
  \label{eq:w-func}
  W(x) &= \bar{N}(x) \left(1 + \int_{x_0}^{x} e^{-K(x')} \eta(x') \, dx' \right).
\end{align}
\rev{Note that $P_0(x) = 0$ when $\eta = 0$, as expected.}

The probability that an avalanche contains $n > 0$ electrons is given by a geometric distribution multiplied by $1 - P_0(x)$
\begin{equation}
  \label{eq:pN}
  P_n(x) = \left[1 - P_0(x)\right] \, P_g(x) \, \left(1-P_g(x)\right)^{n-1},
\end{equation}
with $P_g$ given by
\begin{equation}
  \label{eq:pgeom}
  P_g(x) = \frac{1}{W(x)} = \frac{1 - P_0(x)}{\bar{N}(x)}.
\end{equation}
\rev{Although equation~\eqref{eq:pN} was derived analytically in~\cite{Kendall_1948}, the underlying assumption that $\alpha$ and $\eta$ only depend on $x$ is not always valid, as discussed in~\cite{VanBrunt_1991}.
  Nevertheless, our results in section~\ref{sec:aval-size-distr} will show that a geometric distribution often describes the avalanche size distribution rather well.
}

\subsection{Distribution for number of ionizations in an avalanche}
\label{sec:numb-ioniz-avalanche}

It is important to accurately predict the number of secondary electrons produced by an avalanche.
For processes like ion-induced electron emission and photoionization the number of secondary electrons is approximately proportional to the total number of ionizations $M$ caused by an avalanche.
We will refer to $M$ as the avalanches size.
In~\cite{Kendall_1948}, it was shown that the mean of $M$ is given by
\begin{equation}
  \label{eq:M}
  \bar{M}(x) = 1 + \int_{x_0}^{x} \rev{\bar{N}}(x') \alpha(x') \, dx',
\end{equation}
where the initial ionization was included (hence the $1 + \dots$).
In analogy with $\bar{N}(x) = e^{K(x)}$, we define
\begin{equation}
  \label{eq:kstar}
  K^*(x) = \log\left[\bar{M}(x)\right].
\end{equation}
In attaching gases, $\bar{M}(x)$ can be significantly larger than $\bar{N}(x)$.
For example, if $\alpha(x) = \eta(x) = c$, $\bar{N}(x) = 1$ while $\bar{M}(x) = 1 + c (x - x_0)$.
However, when most of the ionization is produced in a region where $\alpha \gg \eta$ differences will be small and $K^* \approx K$.

No general distribution for $M$ could be obtained in~\cite{Kendall_1948}.
We therefore propose a simple approximation below\rev{, which reduces to the results from section~\ref{sec:electr-aval-num} when there is no attachment}.
Note that the probability that no ionization has occurred up to $x$ is given by
\begin{equation}
  \label{eq:P-M1}
  P'_{1}(x) = P_\mathrm{no}(x) + \int_{x_0}^{x} P_\mathrm{no}(x') \eta(x') \, dx',
\end{equation}
where $P_\mathrm{no}(x')$ is the probability that no ionization or attachment has taken place
\begin{equation}
  \label{eq:P-no}
  P_\mathrm{no}(x) = \exp\left(-\int_{x_0}^{x} [\alpha(x') + \rev{\eta}(x')] \, dx'\right),
\end{equation}
and where the integral in equation~\eqref{eq:P-M1} is the probability that attachment has occurred before any ionization.

We approximate the distribution of $M$ analogous to equation~\eqref{eq:pN}, so that sampling is done as follows.
First, with probability $P'_{1}(x)$ take $M = 1$.
Otherwise, take $M = 1 + m$, with $m$ sampled according to a geometric distribution
\begin{equation}
  \label{eq:P-M}
  P'_{m}(x) = P'_g(x) \, \left(1-P'_g(x)\right)^{m-1},
\end{equation}
where
\begin{equation}
  \label{eq:PMgeom}
  P'_g(x) = \frac{1 - P'_{1}(x)}{\bar{M}(x) - 1}.
\end{equation}
With this choice, the expected value of $M$ is $\bar{M}(x)$.

Note that without attachment we have $\bar{N}(x) = \bar{M}(x)$, $P_0(x) = 0$ and $P'_{1}(x) = 1/\bar{N}(x)$.
This means that $P'_g(x) = 1/\bar{N}(x)$ and that the distributions for $N$ and $M$ are identical.
However, in the more general case with attachment, our distribution for $M$ is an approximation.
Consider a case with $\alpha$ slightly larger than $\eta$ and a long integration domain, so that $M \gg 1$.
Since $\alpha \approx \eta$ we have $P_1' \approx 0.5$.
The probability $P_2'$ of the initial electron producing one additional electron, followed by the loss of both electrons due to attachment, is then on the order of $0.5^3$, much higher than the probability according to equation~\eqref{eq:P-M}.
It would perhaps be possible to add a correction for small avalanche sizes to equation~\eqref{eq:P-M}.
We have not tried to do so, since we expect that undersampling small avalanches will not significantly affect inception probabilities computed by the model.

\subsection{Ionization coefficient for spatial growth}
\label{sec:diffusion-correction}

In the model, we need to use ionization and attachment coefficients that describe the spatial growth of electron avalanches \rev{until they reach a boundary.
  Spatial growth coefficients in general differ from the temporal growth coefficients that are commonly used in plasma fluid models.
  There are two main reasons for this.
  First, an electron avalanche spreads out due to diffusion, so the front of an avalanche will hit a boundary before its tail.
  Second, because ionization is more likely for electrons at the front of an avalanche, the avalanche's center of mass will move with a higher velocity than the electron drift velocity.
  These effects are quantified below, by deriving an approximate formula that describes the spatial growth of an avalanche.
}

Consider an electron avalanche starting from a single electron at $x = 0$ and $t = 0$, which propagates in the $x$-direction in a uniform electric field.
The electron density along the $x$-axis can then be approximated by~\cite{Vass_2017}
\begin{equation}
  \label{eq:ne-xt}
  n_e(x, t) = \frac{1}{\sqrt{4\pi \bar{D}_{L} t}}
  \exp\left(\nu_\mathrm{eff} t - \frac{(x - W t)^2}{4 \bar{D}_{L} t} \right),
\end{equation}
where $\bar{D}_{L}$ is the longitudinal bulk diffusion coefficient, $W$ is the bulk drift velocity, and $\nu_\mathrm{eff}$ is the effective ionization frequency.
Note that bulk coefficients describe evolution of the swarm as a whole, whereas normal (also referred to as `flux') coefficients describe the average properties of individual electrons~\cite{Petrovic_2009}.
For example, $W$ is the velocity of the avalanche's center of mass, which can be higher than the electron drift velocity $v_d$.

\rev{We want to know how many electrons will reach a boundary at some location $x$.
The electron flux at $x$ contains a drift and a diffusive component and is given by}
\begin{equation}
  \label{eq:flux-def}
  \Gamma_e(x, t) = v_d \, n_e(x, t) - D_L \, \partial_x n_e(x, t),
\end{equation}
where $v_d$ is the flux electron drift velocity and $D_L$ the flux longitudinal electron diffusion coefficient.
The total number of electrons passing through $x$ over time is~\cite{Blevin_1984}
\begin{equation}
  \label{eq:ne-x}
  N_e(x) = \int_0^\infty \Gamma_e(x, t) \, dt =
  (c_\mathrm{drift} + c_\mathrm{diff}) \exp(\alpha_T z),
\end{equation}
where we express the coefficients using the dimensionless quantity $\kappa = 4 \bar{D}_{L} \nu_\mathrm{eff}/W^2$ as
\begin{align}
  \label{eq:alpha-T}
  \alpha_T &= \frac{W}{2 \bar{D}_L} \left(1 - \sqrt{1 - \kappa}\right),\\
  \label{eq:c-drift}
  c_\mathrm{drift} &= \frac{v_d}{W} \, \frac{1}{\sqrt{1 - \kappa}},\\
  \label{eq:c-diff}
  c_\mathrm{diff} &= \frac{1}{2} \, \frac{D_L}{\bar{D}_L} \,
                    \left(1 - \frac{1}{\sqrt{1-\kappa}}\right).
\end{align}

The above expression for $\alpha_T$ is used to describe the spatial growth of the electron density in Steady State Townsend experiments, see e.g.~\cite{Blevin_1984,Vass_2017,Hagelaar_2005}.
However, we are not aware of earlier publications in which the expressions for $c_\mathrm{drift}$ and $c_\mathrm{diff}$ were given.
Note that the limit $\kappa \to 0$ corresponds to $\alpha_T = \nu_\mathrm{eff}/W$, that $\alpha_T$ is an effective ionization coefficient, and that for $\kappa > 1$ the integral diverges.

\begin{figure}
  \centering
  \includegraphics[width=\linewidth]{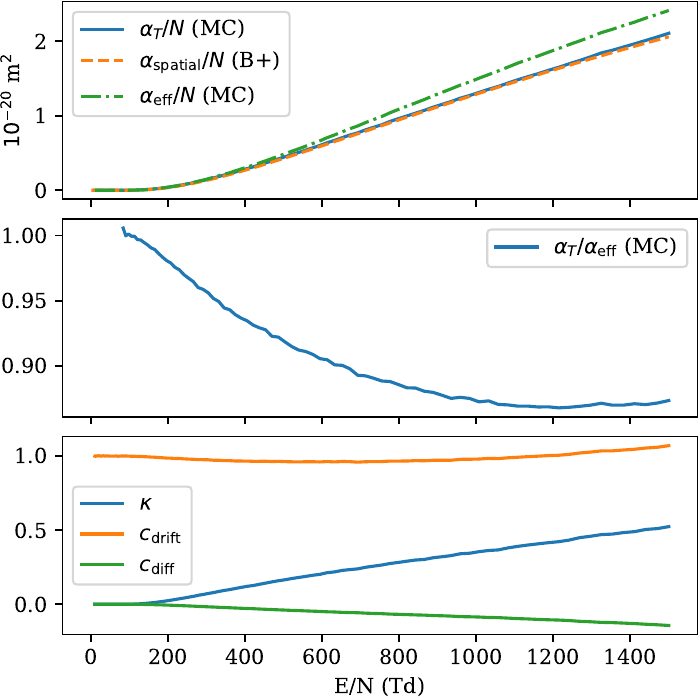}
  \caption{Top: Ionization coefficient in N$_2$, using Biagi cross sections~\cite{Biagi_lxcat}. The curve $\alpha_T/N$ was computed according to equation~\eqref{eq:alpha-T} with a Monte Carlo (MC) swarm code, while $\alpha_\mathrm{spatial}/N$ was computed using BOLSIG+ with a spatial growth model. For comparison, $\alpha_\mathrm{eff}$ corresponding to temporal growth is also shown.
    \rev{Middle: ratio between $\alpha_T$ and $\alpha_\mathrm{eff}$.}
  Bottom: the coefficients $\kappa = 4 \bar{D}_{L} \nu_\mathrm{eff}/W^2$, $c_\mathrm{drift}$ and $c_\mathrm{diff}$ from equations~\eqref{eq:ne-x}--\eqref{eq:c-diff}.}
  \label{fig:alpha-T}
\end{figure}

Figure~\ref{fig:alpha-T} shows a comparison of $\alpha_T$ and $\alpha_\mathrm{eff}$ in N$_2$, with the transport coefficients computed by BOLSIG+~\cite{Hagelaar_2005} and the \texttt{particle\_swarm} Monte Carlo code~\cite{JannisTeunissen_2025}.
The coefficients $c_\mathrm{drift}$ and $c_\mathrm{diff}$ are also shown.
With BOLSIG+, we can directly compute the $\alpha$ and $\eta$ corresponding to the $\alpha_T$ from equation~\eqref{eq:alpha-T} by using a spatial growth model.
With the \texttt{particle\_swarm} code, we determine $\alpha_T$ according to equation~\eqref{eq:alpha-T} and obtain spatial \rev{ionization and attachment} coefficients as
\begin{align}
  \label{eq:alpha-eta-corrected}
  \alpha_\mathrm{spatial} &= \alpha \, \left(\alpha_T / \alpha_\mathrm{eff}\right),\\
  \eta_\mathrm{spatial} &= \eta \, \left(\alpha_T / \alpha_\mathrm{eff}\right),
\end{align}
\rev{where $\alpha_{\mathrm{eff}} = \alpha - \eta$.}
There is generally good agreement between the BOLSIG+ and Monte Carlo (MC) results, but the relative difference between the curves exceeds 10\% at relatively low fields, as discussed in~\cite{Hagelaar_2025}.

Note that $c_\mathrm{drift}$ is close to one, while $c_\mathrm{diff}$ is close to zero except for very high fields.
Furthermore, the diffusive term in equation~\eqref{eq:flux-def} is not correct when a solid wall is placed at $x$, since there can be diffusion towards the wall but not from it.
We will therefore assume that the prefactor $c_\mathrm{drift} + c_\mathrm{diff}$ in equation~\eqref{eq:ne-x} is one.
With this assumption, no correction is required to the expressions from sections~\ref{sec:electr-aval-num} and \ref{sec:numb-ioniz-avalanche}, except for the use of spatial growth coefficients.

\subsection{Multi-avalanche generation}
\label{sec:multi-aval-gener}

We assume that avalanches propagate with a velocity $-\mu_e(E) \mathbf{E}$, where $\mu_e(E)$ is the electron mobility and $\mathbf{E}$ is the electric field.
For consistency with equation~\eqref{eq:ne-xt} $\mu_e$ should be the bulk mobility, but using the flux mobility instead will only affect the avalanche travel time defined below.
Given any initial electron position in the domain, we can trace the avalanche path and the travel time along the path.
The integrals from section~\ref{sec:numb-ioniz-avalanche} can be computed simultaneously, although we actually precompute them as explained in section~\ref{sec:avalanche-sim-method}.
We define the arrival position of the avalanche $\mathbf{r}_\mathrm{arrival}$ as the location along the path at which the avalanche size $\bar{M}-1$ is half of its total value $\bar{M}_\mathrm{total} - 1$ as given by equation~\eqref{eq:M}:
\begin{equation}
  \label{eq:r-arrival}
  \bar{M}(\mathbf{r}_\mathrm{arrival}) - 1 = \tfrac{1}{2} \left(\bar{M}_\mathrm{total} - 1\right).
\end{equation}
One is subtracted from both sides since we do not account for the initial ionization.
The avalanche travel time $t_\mathrm{travel}$ is defined as the time corresponding to $\mathbf{r}_\mathrm{arrival}$.
Furthermore, the arrival time of an avalanche created at $t_\mathrm{creation}$ is
\begin{equation}
  \label{eq:t-arrival}
  t_\mathrm{arrival} = t_\mathrm{creation} + t_\mathrm{travel}.
\end{equation}

An initial avalanche can generate additional avalanches due to secondary electron emission, as described in sections~\ref{sec:photoionization}--\ref{sec:see-pos-ions}.
We use a discrete-event simulation technique, keeping track of all future avalanches in a list.
For every avalanche $t_\mathrm{arrival}$ is determined.
At every step, time advances to the smallest $t_\mathrm{arrival}$ and the corresponding avalanches is taken from the list.
The size of this avalanche is sampled, and if secondary emission is produced, the sizes of the resulting avalanches are sampled.
Secondary avalanches that produce additional ionization are added to the list of future avalanches.

\subsection{Inception criterion}
\label{sec:inception-criterion}

Whether discharge inception will take place is determined according to two criteria.
First, if the number of future avalanches $n_\mathrm{future}$ in the event list exceeds a threshold
\begin{equation}
  \label{eq:num-future-av}
  n_\mathrm{future} \geq n_\mathrm{inc},
\end{equation}
this indicates self-sustained growth that will cause a discharge.
Second, if an avalanche's size $M$ exceeds a threshold
\begin{equation}
  \label{eq:av-size-threshold}
  M \geq M_\mathrm{inc},
\end{equation}
it is assumed that an avalanche-to-streamer transition has occurred.
We here use $n_\mathrm{inc} = 10^3$ and $M_\mathrm{inc} = 10^8$ unless specified otherwise.
When inception occurs, the inception time is stored.
If inception has not occurred and there are no more future avalanches, there is no inception.

The probability of inception $p_\mathrm{inc}(\mathbf{r})$ is estimated by starting $N_\mathrm{runs}$ avalanches from $\mathbf{r}$ and determining the fraction of runs that lead to inception.
The mean inception time is determined as the average of the inception times, only considering the runs with inception.
Furthermore, a spatially-averaged inception probability $\bar{p}_\mathrm{inc}$ is computed as
\begin{equation}
  \label{eq:avg-p-inception}
  \bar{p}_\mathrm{inc} = (1/V_\mathrm{gas}) \, \int_\mathrm{gas} p_\mathrm{inc}(\mathbf{r}) \, dV,
\end{equation}
where $V_\mathrm{gas}$ is the gas volume in the domain.

\subsection{Photoionization and photoemission}
\label{sec:photoionization}

Photoionization is an important process for discharges in air~\cite{Nijdam_2010}, in which an oxygen molecule is ionized by a photon emitted from an excited nitrogen molecule.
Our implementation of photoionization is based on the stochastic variants of Zheleznyak's model~\cite{Zheleznyak_1982} used in e.g.~\cite{Chanrion_2008,Teunissen_2016,Marskar_2020}.
\rev{Photons with wavelengths between $\lambda_\mathrm{min} = 98.0 \, \mathrm{nm}$ and $\lambda_\mathrm{max} = 102.5 \, \mathrm{nm}$ are considered.}
The number of ionizing photons is sampled from a Poisson distribution with mean
\begin{equation}
  \label{eq:n-photons}
  N_\gamma = \xi \frac{p_q}{p + p_q} M,
\end{equation}
where $p_q$ is the quenching pressure, $\xi$ is the photoionization efficiency, and $M$ is the avalanche size.
For simplicity, all ionizing photons are assumed to originate from a point source $\mathbf{r}_\mathrm{source}$ at $\mathbf{r}_\mathrm{arrival}$.

For each photon a random isotropic unit vector $\hat{\mathbf{n}}$ is sampled.
\rev{
  Photon absorption coefficients $\lambda$ are sampled as
\begin{equation}
  \label{eq:photo-abs-length}
  \lambda = (\chi_\mathrm{min} \, p_{\mathrm{O}_2})^R \, (\chi_\mathrm{max} \, p_{\mathrm{O}_2})^{1-R},
\end{equation}
where $R$ is a uniform random number and $\chi_\mathrm{min}$ and $\chi_\mathrm{max}$ are the minimum and maximum absorption coefficients.}
The actual photon absorption distance $d$ is sampled from the exponential distribution with mean \rev{$\lambda^{-1}$}, so that the absorption location (in the absence of surfaces) is given by
\begin{equation}
  \label{eq:photo-abs-pos}
  \mathbf{r}_\mathrm{absorption} = \mathbf{r}_\mathrm{source} + d \, \hat{\mathbf{n}}.
\end{equation}
For dry air, we here use the following coefficients~\cite{Zheleznyak_1982}: $p_q = 40 \, \textrm{mbar}$, $\xi = 0.075$, \rev{$\chi_\mathrm{min} = 2.625 \times 10^{3} \, \mathrm{m}^{-1} \, \mathrm{bar}^{-1}$ and $\chi_\mathrm{max} =  1.500 \times 10^5 \, \mathrm{m}^{-1} \, \mathrm{bar}^{-1}$}.

Photon paths can be traced through the numerical mesh to detect when they hit a surface, which is enabled for all the examples shown in this paper.
If a photon hits a surface, a new electron is generated at the surface with a probability $\gamma_\mathrm{surf}$ that can be specified per material in the input.
If photon path tracing is disabled, or no surface is hit, an electron is generated at $r_\mathrm{absorption}$ as long as this position is inside the gas.

Sometimes, a single avalanche can produce so many secondary avalanches that the inception criterion of equation~\eqref{eq:num-future-av} is met.
This does not always indicate sustained growth in the number of avalanches over time, for example if the primary avalanche started from a location where secondary electron emission is very unlikely.
We therefore limit the number of photon-produced secondary avalanches by a single primary avalanche to $n_\mathrm{inc}/2$.
The effect of this limit is illustrated in section~\ref{sec:incept-prob-plates}.





\subsection{Secondary emission from positive ions}
\label{sec:see-pos-ions}

Positive ions are assumed to propagate with a velocity $\mu_i \mathbf{E}$.
For an avalanche of size $M$ arriving at $\mathbf{r}_\mathrm{arrival}$, the field is traced until a domain boundary or a material other than gas is reached at $\mathbf{r}_\mathrm{boundary}$, and the boundary material is stored.
The ion travel time is also determined.

An ion secondary emission coefficient $\gamma_\mathrm{ion}$ can be specified for each boundary material.
For simplicity, it is assumed that there is a single positive ion species.
The production and conversion of ions to different types are thus not taken into account.
The number of secondary electrons produced at $\mathbf{r}_\mathrm{boundary}$ is sampled from a Poisson distribution with mean $\gamma_\mathrm{ion} M$.

\subsection{Determining the inception voltage}
\label{sec:inception-voltage-method}

It is often important to determine the partial discharge inception voltage (PDIV).
In a probabilistic model, a small but finite threshold $p_\mathrm{threshold}$ for the spatially-averaged inception probability $\bar{p}_\mathrm{inc}$ first has to be defined, and then the voltage corresponding to this threshold can be determined.
We assume all dielectrics to be linear, so that multiplying the applied voltage with a factor $f_\mathrm{field}$ is equivalent to multiplying the electrostatic field by a factor $f_\mathrm{field}$.

A Python script is provided that automatically determines the PDIV\@.
As input, a lower and upper bound for $f_\mathrm{field}$ and a value for $p_\mathrm{threshold}$ have to be specified.
First, the voltage is varied linearly from low to high, to determine more narrow bounds for $f_\mathrm{field}$ at which $g(f_\mathrm{field}) = \bar{p}_\mathrm{inc} - p_\mathrm{threshold}$ changes sign.
Then bisection is performed to further narrow down the interval at which $g(f_\mathrm{field})$ has a root, with an early exit when the stochastic function $g(f_\mathrm{field})$ shows non-monotonic behavior.
Finally, the classic Robbins-Monro algorithm~\cite{Robbins_1951} is used to refine the root using a user-defined number of iterations.

\subsection{Particle simulations to test avalanche model}
\label{sec:particle-sim}

Our code includes a standard particle model for electrons~\cite{Teunissen_2016}, which can be used to test the approximations made for electron avalanches.
With the particle model, individual electrons are tracked as particles that stochastically collide with gas molecules, using the null-collision method.
The model requires electron-neutral cross sections as input.
At every time step, the acceleration of the electrons is updated by interpolating the electric field at their current position, and particles are removed when they leave the gas.
Photoionization can be included, but ion secondary emission mechanisms are not implemented, since particle simulations on ion time scales can be computationally very costly.

\section{Implementation}
\label{sec:implementation}

Below, we discuss the main implementation aspects that relate to functionality or performance.

\subsection{Unstructured grid format}
\label{sec:ugrid-format}

Several types of unstructured grids can be provided as input, which are converted to a binary format using the \texttt{meshio} library~\cite{Schlomer_2024}.
Afterwards, the neighbors of each cell are determined, and normal vectors of all cell faces are computed.
Three types of cells are supported: triangles and quadrilaterals in 2D and tetrahedra in 3D.
The components of the electric field have to be defined at vertices of the grid.
A parameter describing the material type can either be stored at vertices or as a cell variable.
By convention, materials are indicated by integer values starting from zero, with zero corresponding to the gas.

\subsection{Fast interpolation and integration}
\label{sec:ugrid-interp}

We have implemented efficient linear interpolation on unstructured grids, using barycentric coordinates.
To interpolate at some position $\mathbf{r}$, the cell containing $\mathbf{r}$ first has to be found.
If the index of a nearby cell is already known, we use the `particle-localization algorithm' described in~\cite{Haselbacher_2007}.
Otherwise, a $k$-d tree search~\cite{Kennel_2004} is performed to localize a nearby cell.

For tracing the electric field and computing the integrals given by equations~\eqref{eq:K-func} and~\eqref{eq:P-no} we use the cell-traversal method from~\cite{Haselbacher_2007} combined with the Bogacki–Shampine Runge-Kutta method~\cite{Bogacki_1989} with adaptive step size control.
For the results presented here we use a relative and absolute tolerance of $10^{-5}$ per step.
Using the samples along the path, the other integrals described in section~\ref{sec:numb-ioniz-avalanche} are computed.
Special care has to be taken of integrals in which an exponential term rapidly varies, as discussed in~\ref{sec:appendix-exp-int}.

\subsection{Avalanche simulation}
\label{sec:avalanche-sim-method}

Most parameters required to simulate avalanches are precomputed on the vertices of the unstructured grid:
\begin{itemize}
  \item $P'_{1}$ and $K^*$ are used to sample the avalanche size, see section~\ref{sec:numb-ioniz-avalanche}.
  \item The avalanche `arrival' location $\mathbf{r}_\mathrm{arrival}$ and travel time $t_\mathrm{travel}$.
  \item The location where positive ions hit a surface $\mathbf{r}_\mathrm{boundary}$ and the ion secondary emission coefficient $\gamma_\mathrm{ion}$ of the surface.
\end{itemize}
For an initial electron at a position $\mathbf{r}$, these parameters are linearly interpolated to $\mathbf{r}$.
With a probability $1 - P'_{1}$ a new avalanche is stored, together with the above parameters and its arrival time.
When this avalanche is created, the stored parameters can be used to efficiently sample its size and its secondary emission.

A priority queue is used to efficiently select the avalanche with the lowest arrival time.
Avalanches starting from different locations are simulated in parallel using OpenMP.

\subsection{Computing the inception probability}
\label{sec:inception-prob-method}

Our model can compute the probability of discharge inception $p_\mathrm{inc}(\mathbf{r})$ for a single initial electron appearing at $\mathbf{r}$.
In principle, this can be done by first performing $N_\mathrm{runs}$ simulations with a first electron at $\mathbf{r}$, and determining $p_\mathrm{inc}(\mathbf{r})$ as the fraction of runs with inception.
However, it is possible to reduce the variance of such a measurement.

First, we determine the expected number of avalanches as $\bar{N}_\mathrm{av} = N_\mathrm{runs} \, (1 - P'_{1})$, not counting the `avalanches' that produce no further ionization.
Then $N^*_\mathrm{runs} = \lceil \bar{N}_\mathrm{av}\rceil$ runs are performed, and the probability of inception is determined as the number of runs with inception multiplied by a correction factor $\bar{N}_\mathrm{av} / N^*_\mathrm{runs}$.

Second, we use \rev{a simple variant of so-called antithetic variates~\cite{hammersleyNewMonteCarlo1956}}.
For the first half of the runs, uniform $[0, 1)$ random numbers $u_1, u_2, \dots$ are used to sample the size of the initial avalanche.
Specifically, we draw samples from a geometric distribution with probability $p$ as
\begin{equation}
  \label{eq:sample-geom}
  X = \log(1 - u) / \log(1 - p),
\end{equation}
see equation~\eqref{eq:PMgeom}.
(We remark that the actual implementation is more complicated than this expression, to avoid numerical issues for $u$ close to zero or one in $\log(1 - u)$.)
For the second half of the runs, the complement \rev{$1-u_1, 1-u_2, \dots$ of the random numbers} is used to determine the size of the first avalanche.
With this procedure, the initial avalanche sizes are better balanced, reducing the variance of the inception probability\rev{ estimate}.

\subsection{Transport data}
\label{sec:transport-data}

For electrons, the code needs to know the ionization coefficient $\alpha(E)$, attachment coefficient $\eta(E)$ and mobility $\mu_e(E)$ as function of the electric field strength $E$.
We compute such data with a Boltzmann solver such as BOLSIG+~\cite{Hagelaar_2005} or \texttt{particle\_swarm}~\cite{JannisTeunissen_2025} from electron-neutral cross sections obtained from the LXCAT~\cite{Carbone_2021,Pitchford_2017,Pancheshnyi_2012} website.
The resulting reduced transport coefficients are read in by the code, converted to non-reduced units, and stored in a lookup table for fast interpolation.
The same input data can thus be used when the gas number density changes, as long as the gas composition stays the same.
If during a simulation electric fields are encountered that exceed the range of the input data, linear extrapolation is performed using the slope of the last two data points.

For gas mixtures including O$_2$, care must be taken to account for three-body attachment
\begin{equation}
  \label{eq:three-body-O2}
  e + \mathrm{O}_2 + \mathrm{M} \to \mathrm{O}_2^- + \mathrm{M},
\end{equation}
where the third body can for example be O$_2$ or H$_2$O~\cite{Aleksandrov_2025}.
We include the three-body cross section in the input of the Boltzmann solver to obtain a rate coefficient $k_3(E)$ for the process, in units of $\mathrm{m}^6/\mathrm{s}$.
A corresponding three-body attachment coefficient is then computed as
\begin{equation}
  \label{eq:three-body-eta}
  \eta_3(E) = \frac{k_3(E)}{\mu_e(E) E} \, [\mathrm{O}_2] \,
  \left( c_1 [\mathrm{M}_1] + c_2 [\mathrm{M}_2] + \dots \right),
\end{equation}
where $[\mathrm{O}_2]$ is the number density of O$_2$ molecules.
By default we take into account O$_2$ (with a coefficient of 1.0) and H$_2$O (with a coefficient of 6.0) as third bodies~\cite{Aleksandrov_2025}, but this can be customized in the input.
The attachment coefficient is then set to the sum of $\eta_3(E)$ and the two-body attachment coefficient computed by the Boltzmann solver.

The model does not distinguish between different positive ion species.
The ion mobility is described by a constant reduced mobility coefficient $\mu_i \, N$, where $N$ is the gas number density.
For the results in this paper we \rev{have used $\mu_i \, N = 4.8 \times 10^{21} \, \mathrm{m^{-1} V^{-1} s^{-1}}$, which corresponds to an ion mobility of $2.0 \times 10^{-4} \, \textrm{m}^2 \textrm{V}^{-1} \textrm{s}^{-1}$ at $1 \, \textrm{bar}$ and $300 \, \textrm{K}$.
   This value of $\mu_i\, N$ was motivated by the mobilities for $\mathrm{N}_4^+$ and $\mathrm{N}_2^+$ in $\mathrm{N}_2$ and in air reported in~\cite{ellisTransportPropertiesGaseous1978,viehlandTransportPropertiesGaseous1995,benhenniAnalysisIonMobility2006}.
In our model the ion mobility is only} used to estimate the travel time of positive ions until they hit a boundary, where they can possibly generate secondary electron emission.

\subsection{Output}
\label{sec:output}

Output is written to the same unstructured grid as used for the input, in the VTK unstructured format.
The output grid contains all the avalanche parameters described in section~\ref{sec:avalanche-sim-method}, and several other parameters such as $\alpha$, $\eta$ and $K$.
What else is stored depends on the type of computation that is performed:
\begin{itemize}
  \item If only the avalanche parameters are computed, no additional output is stored.
  \item If avalanches are simulated the output also contains the estimated inception probability and inception time. Furthermore, the volume-averaged inception probability $\bar{p}_\mathrm{inc}$, an estimate of the standard deviation of $\bar{p}_\mathrm{inc}$, and the volume of the gas $V_\mathrm{gas}$ are stored.
  \item If a particle simulation is performed, the output contains a text file with the avalanche sizes for a given initial electron location, for a specified number of runs.
\end{itemize}

\section{Results}
\label{sec:results}

\subsection{Avalanche size distribution}
\label{sec:aval-size-distr}

We can compare the avalanche size distribution given in section~\ref{sec:numb-ioniz-avalanche} against particle simulations.
We remind the reader that avalanche size here refers to the number of ionizations ($M$) produced by the avalanche, and not to the number of electrons $N$ in the avalanche at some time.
As a first test, we consider a homogeneous electric field $\mathbf{E}_0 = E_0 \, \hat{\mathbf{z}}$, with an initial electron at $\mathbf{r}_0 = (0, 0, z_0)$ moving towards a plate electrode at $z = z_0 - d$.
Three gases are considered, all at 1\,bar and 300\,K: pure N$_2$, artificial air (80\% N$_2$, 20\% O$_2$), and 96\% N$_2$ with 4\% SF$_6$.
Transport data is computed with the \texttt{particle\_swarm} code using Siglo cross sections for N$_2$~\cite{Phelps_1985,Phelps_lxcat}, MuroranIT cross sections for O$_2$~\cite{Kawaguchi_2025,MuroranIT_lxcat} and Biagi cross sections for SF$_6$~\cite{Biagi_lxcat}.

\begin{figure}
  \centering
  \includegraphics[width=\linewidth]{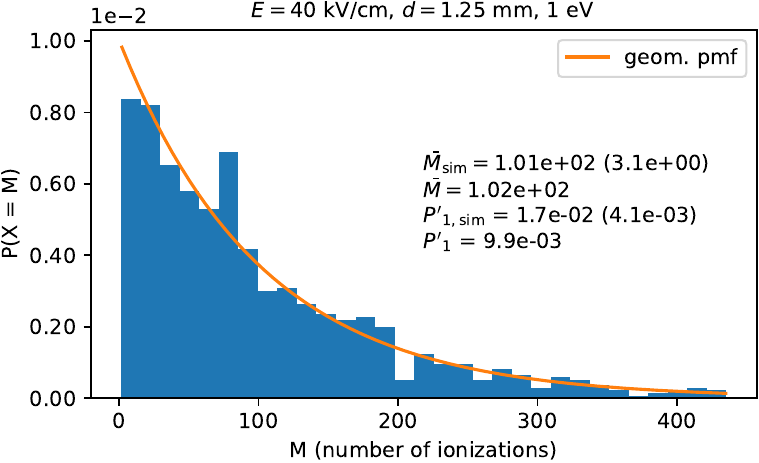}\\[0.2em]
  \includegraphics[width=\linewidth]{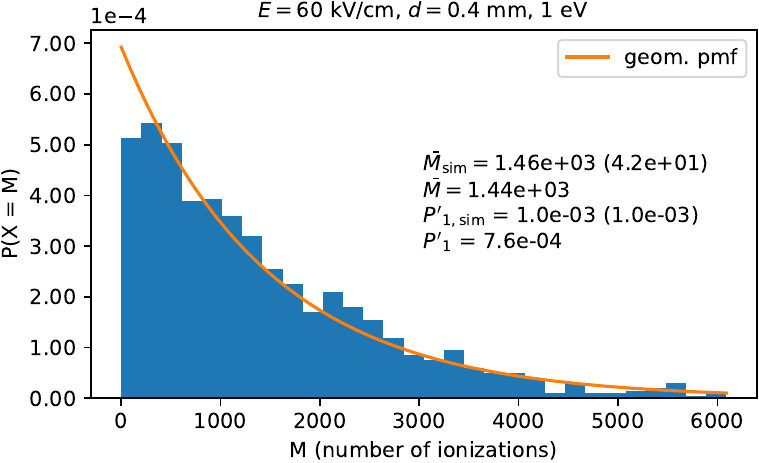}\\[0.2em]
  \includegraphics[width=\linewidth]{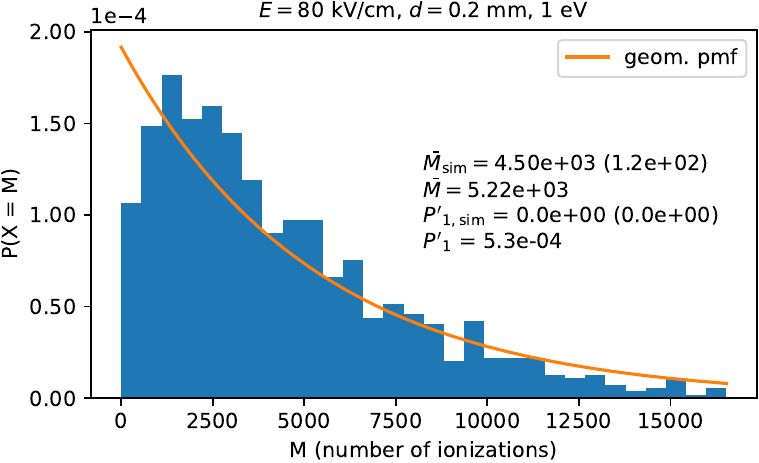}
  \caption{Avalanche size distributions in uniform electric fields in N$_2$, for various gap sizes $d$, for an initial electron energy of $1 \, \textrm{eV}$.
    The bars indicate results from 1000 particle simulations, and the curve shows the probability mass function according to equation~\eqref{eq:P-M}.
    The probability $P_1'$ (producing no additional ionization, see equation~\eqref{eq:P-M1}) is given, and also estimated from the simulations.
    This is also done for the mean number of ionizations ($\bar{M}$), see equation~\eqref{eq:M}. Estimated standard deviations are indicated between parentheses.}
  \label{fig:uniform-N2}
\end{figure}

Figure~\ref{fig:uniform-N2} shows results in N$_2$ for several values of $d$ and $E_0$.
In the particle simulations, the velocity of the initial electron was sampled from a Maxwellian distribution corresponding to an energy of $1 \, \textrm{eV}$.
Due to the exponential growth of avalanches, their final size is sensitive to the input data used, and to a lesser extent to the energy of the initial electron.
Taking this into account, there is very good agreement between the theoretical approximations of section~\ref{sec:numb-ioniz-avalanche} and the observed avalanche size distributions.

\begin{figure}
  \centering
  \includegraphics[width=\linewidth]{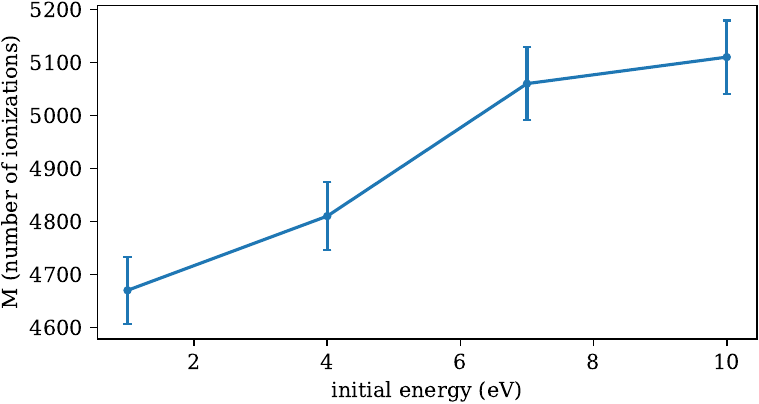}
  \caption{Effect of initial electron energy on the avalanche size, determined using 4000 particle simulations per initial energy.
    The conditions correspond to the bottom case of figure~\ref{fig:uniform-N2} (N$_2$, $E = 80 \, \textrm{kV/cm}$, $d = 0.2 \, \textrm{mm}$).
    The error bars indicate $\pm$ one standard deviation.}
  \label{fig:uniform-N2-energy-effect}
\end{figure}

For the highest field of $80 \, \textrm{kV/cm}$, the avalanches in the particle simulations are about 15\% smaller, which is related to the initial electron energy of 1 eV\@.
If we instead use an initial energy of $7 \, \textrm{eV}$, which is approximately the mean energy in such a field, the mean avalanche size \rev{agrees very well with} equation~\eqref{eq:M}, as illustrated in figure~\ref{fig:uniform-N2-energy-effect}.
In principle, it would be possible to correct for the initial electron energy in equation~\eqref{eq:M}, by taking into account an energy relaxation length.
However, since in a real experiment the energy of the initial electron is unknown, and since uncertainty in the input data can have a much more significant effect on the avalanche size, we do not perform such a correction.

Another effect visible at $80 \, \textrm{kV/cm}$ is that there are fewer small-size avalanches in the particle simulations than predicted by the geometric distribution of equation~\eqref{eq:P-M}.
The discrepancy is caused by the assumption that $\alpha$ depends only on the local electric field strength.
In reality, the probability of ionization depends on the electron energy, which is correlated with previous ionization events caused by an electron.
In high fields, impact ionization is a significant energy loss mechanism for electrons, so an electron that has not caused ionization over some distance will be more likely to do so, and vice versa.
This effect has been studied by several authors, see e.g.~\cite{VanBrunt_1991,Byrne_1969,Kunhardt_1986}.

\begin{figure}
  \centering
  \includegraphics[width=\linewidth]{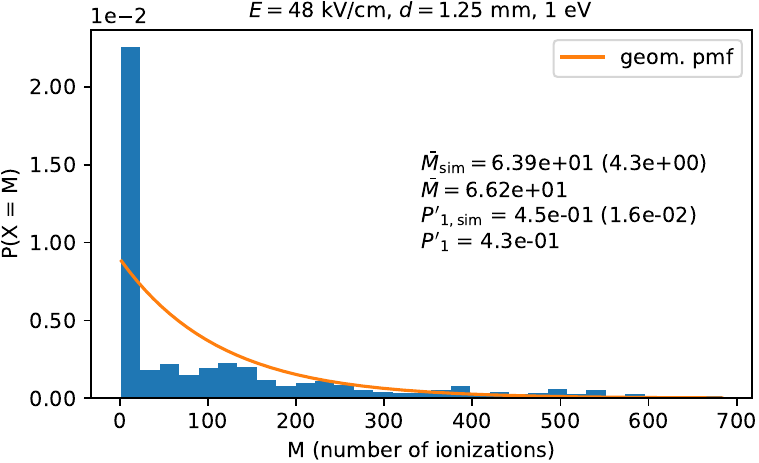}\\[0.2em]
  \includegraphics[width=\linewidth]{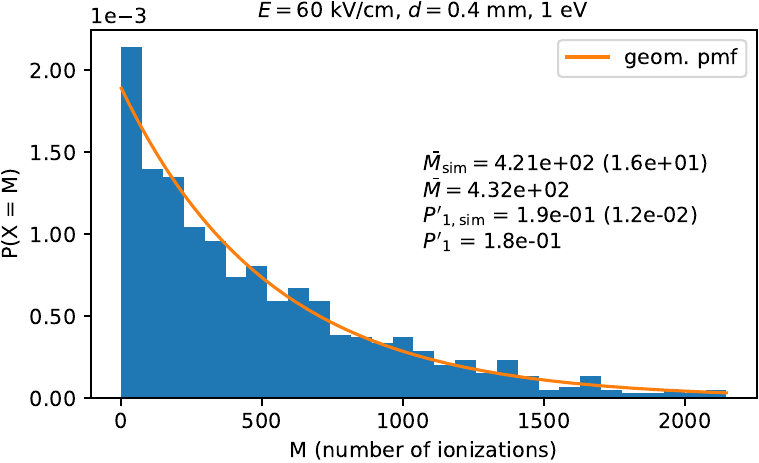}
  \caption{Avalanche size distribution in uniform electric field in N$_2$ containing 4\% SF$_6$. The discrepancy for small avalanche sizes is expected, see the end of section~\ref{sec:numb-ioniz-avalanche}.}
  \label{fig:uniform-sf6}
\end{figure}

In electronegative gases the initial electron can be lost due to attachment, which increases the probability $P_1'$ of producing no further ionization.
Figure~\ref{fig:uniform-sf6} shows results in a uniform field in 96\% N$_2$ and 4\% SF$_6$.
In a field of $48 \, \textrm{kV/cm}$ the geometric distribution underestimates the probability of $M$ (the number of ionizations) being small.
This behavior is expected, as discussed at the end of section~\ref{sec:numb-ioniz-avalanche}.
In a higher field of $60 \, \textrm{kV/cm}$ this difference is less visible due to the smaller value of $P_1'$ and due to the larger bin size of the histogram.
Although small avalanche sizes are undersampled by our approximation, there is good agreement between the particle simulations and the predictions for $\bar{M}$ and $P_1'$.
For the $48 \, \textrm{kV/cm}$ case $K^*$ from equation~\eqref{eq:kstar} is about 23\% larger than $K$ from equation~\eqref{eq:K-func}, whereas for the $60 \, \textrm{kV/cm}$ this difference is less than 1\%.

\begin{figure}
  \centering
  \includegraphics[width=\linewidth]{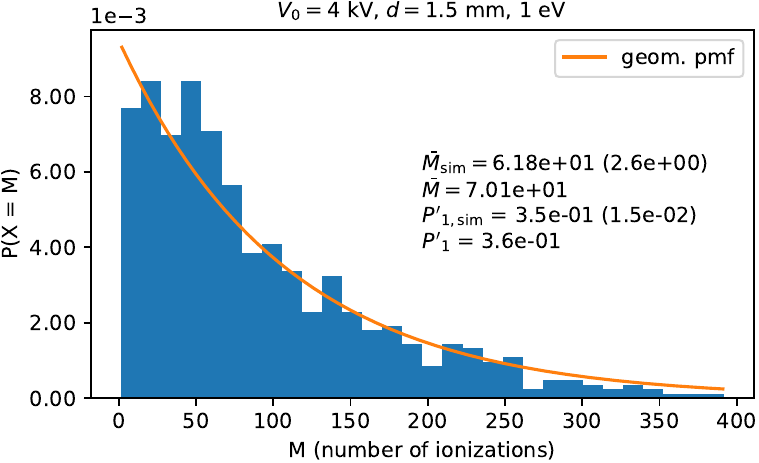}\\[0.2em]
  \includegraphics[width=\linewidth]{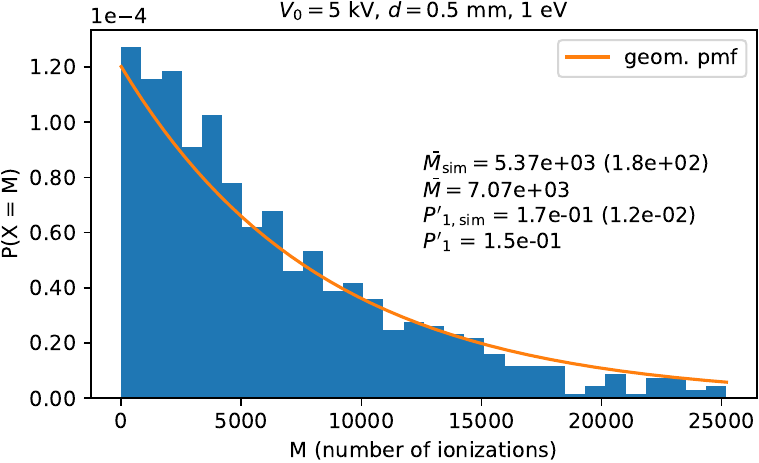}
  \caption{Avalanche size distribution for a conducting sphere of radius $R = 0.5 \, \textrm{mm}$ at a voltage $V_0$ in air (80\% N$_2$, 20\% O$_2$). Initial electrons are placed a distance $d$ from the sphere.}
  \label{fig:sphere-air}
\end{figure}

Next, we consider a non-uniform field, namely that of a conducting sphere with a radius $R = 0.5 \, \textrm{mm}$ at a positive voltage $V_0$ in free space, so that the field only has a radial component $E_r(r) = V_0 \, R \, r^{-2}$ with $r$ the distance from the sphere's center.
Figure~\ref{fig:sphere-air} shows results for two cases in artificial air.
The distribution of avalanche sizes shows good agreement, but the mean avalanche size is somewhat smaller in the particle simulations.
The main reason for this discrepancy is that the local field approximation can be inaccurate in a spatially varying field, since it does not take into account the finite energy relaxation time of electrons.

\subsection{Inception between parallel plates}
\label{sec:incept-prob-plates}

We now determine the probability of discharge inception in a parallel plate geometry, according to the inception definition of section~\ref{sec:inception-criterion}.
The 2D Cartesian geometry consists of two planar electrodes with a width of $2.5 \, \textrm{mm}$, separated by $2.5 \, \textrm{mm}$.
The bottom plate is grounded, while a positive voltage $V_0$ is applied to the top plate.
This results in a homogeneous electric field $E_0$ pointing downwards.
A dielectric material is present on the sides of the domain.
The gas between the electrodes is artificial air (80\% N$_2$, 20\% O$_2$) at 1 bar and 300 K\@.
Photoionization is included as a secondary electron mechanism, using the default parameters given in section~\ref{sec:photoionization}.
Photoemission or ion secondary emission is not included.

\begin{figure}
  \centering
  \includegraphics[width=\linewidth]{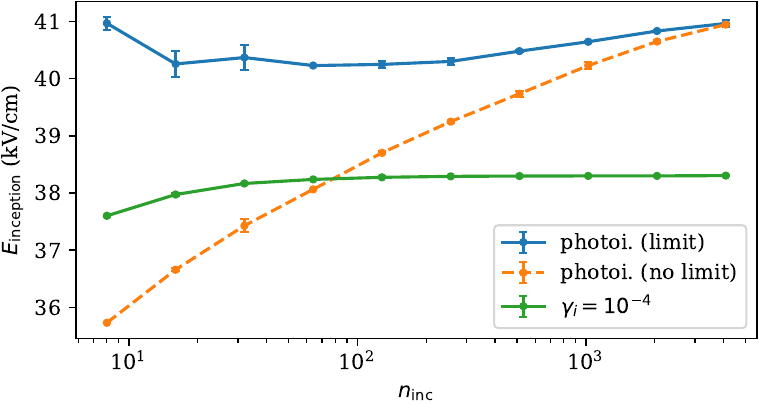}
  \caption{Inception electric field $E_\mathrm{inc} = V_\mathrm{inc}/d$ as a function of the inception threshold $n_\mathrm{inc}$ used in equation~\eqref{eq:num-future-av}.
    Three cases are shown: one in which the number of photoionization events produced by a single avalanche is limited to $n_\mathrm{inc}/2$ (see section~\ref{sec:photoionization}), one in which this number is not limited, and one without photoionization but with an ion secondary emission coefficient $\gamma_i = 10^{-4}$.
    The simulations were performed in air (80\% N$_2$, 20\% O$_2$), in a plate-plate geometry with a distance $d = 2.5\,\mathrm{mm}$ between the plates.}
  \label{fig:effect-ninc}
\end{figure}

We can compute the inception voltage $V_\mathrm{inc}$ corresponding to a certain volume-averaged inception probability $\bar{p}_\mathrm{inc}$, as discussed in section~\ref{sec:inception-voltage-method}.
In a uniform gap with secondary emission, equation~\eqref{eq:num-future-av} is the most relevant inception criterion.
By default, we use $n_\mathrm{inc} = 1000$ as the threshold for the number of future avalanches.
The smaller $n_\mathrm{inc}$ is, the more likely it becomes that inception is `detected' due to stochastic fluctuations, even though eventually all avalanches are expected to disappear.
Figure~\ref{fig:effect-ninc} shows how the inception field $E_\mathrm{inc} = V_\mathrm{inc}/d$ depends on $n_\mathrm{inc}$ for $\bar{p}_\mathrm{inc} = 10^{-4}$.
Three cases are shown: one in which the number of photoionization events produced by a single avalanche is limited to $n_\mathrm{inc}/2$ (see section~\ref{sec:photoionization}), one in which this number is not limited, and one without photoionization but with an ion secondary emission coefficient $\gamma_i = 10^{-4}$.
For the case with $\gamma_i = 10^{-4}$ the inception voltage hardly depends on $n_\mathrm{inc}$ as long as $n_\mathrm{inc} \geq 32$.
On the other hand, if the number of photoionization events is not limited, the inception voltage clearly decreases when $n_\mathrm{inc}$ is reduced.
This happens because a single avalanche starting close to the cathode can then create enough secondary avalanches to meet the inception criterion, even though the number of avalanches is expected to decrease over time.
For the case in which the number of photons is limited, the dependence of the inception voltage on $n_\mathrm{inc}$ is much weaker, with less than 2\% variation.

\begin{figure}
  \centering
  \includegraphics[width=\linewidth]{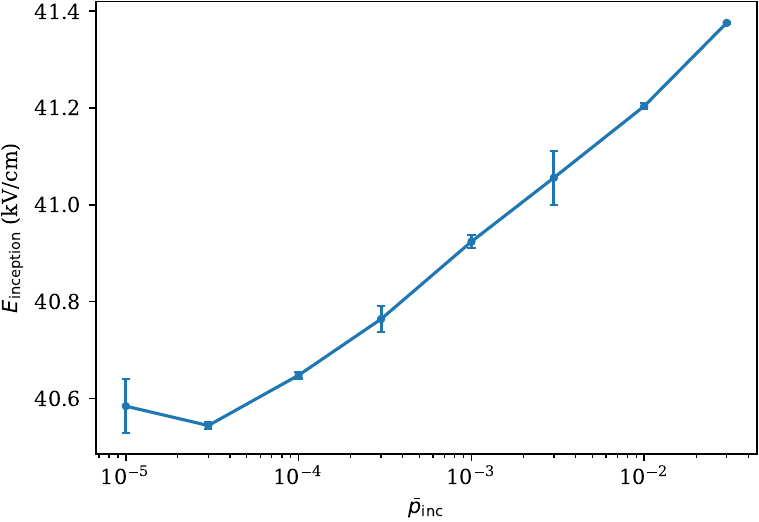}
  \caption{Inception electric field $E_\mathrm{inc} = V_\mathrm{inc}/d$ as a function of the volume-averaged inception probability $\bar{p}_\mathrm{inc}$ in air (80\% N$_2$, 20\% O$_2$), in a plate-plate geometry with a distance $d = 2.5\,\mathrm{mm}$ between the plates. The error bars indicate estimated errors from the optimization procedure described in section~\ref{sec:inception-voltage-method}.}
  \label{fig:pdiv-air-uniform}
\end{figure}

The inception voltage not only depends on $n_\mathrm{inc}$, but also on the threshold used for $\bar{p}_\mathrm{inc}$.
This dependence is illustrated in figure~\ref{fig:pdiv-air-uniform}.
There is about $2\%$ difference in $E_\mathrm{inc}$ when $\bar{p}_\mathrm{inc}$ is varied between $10^{-5}$ and $3\times 10^{-2}$.
However, for small values of $\bar{p}_\mathrm{inc}$ the curve convergences to a value between $40.5 \, \textrm{kV/cm}$ and $40.6 \, \textrm{kV/cm}$.

\begin{figure}
  \centering
  \includegraphics[width=0.8\linewidth]{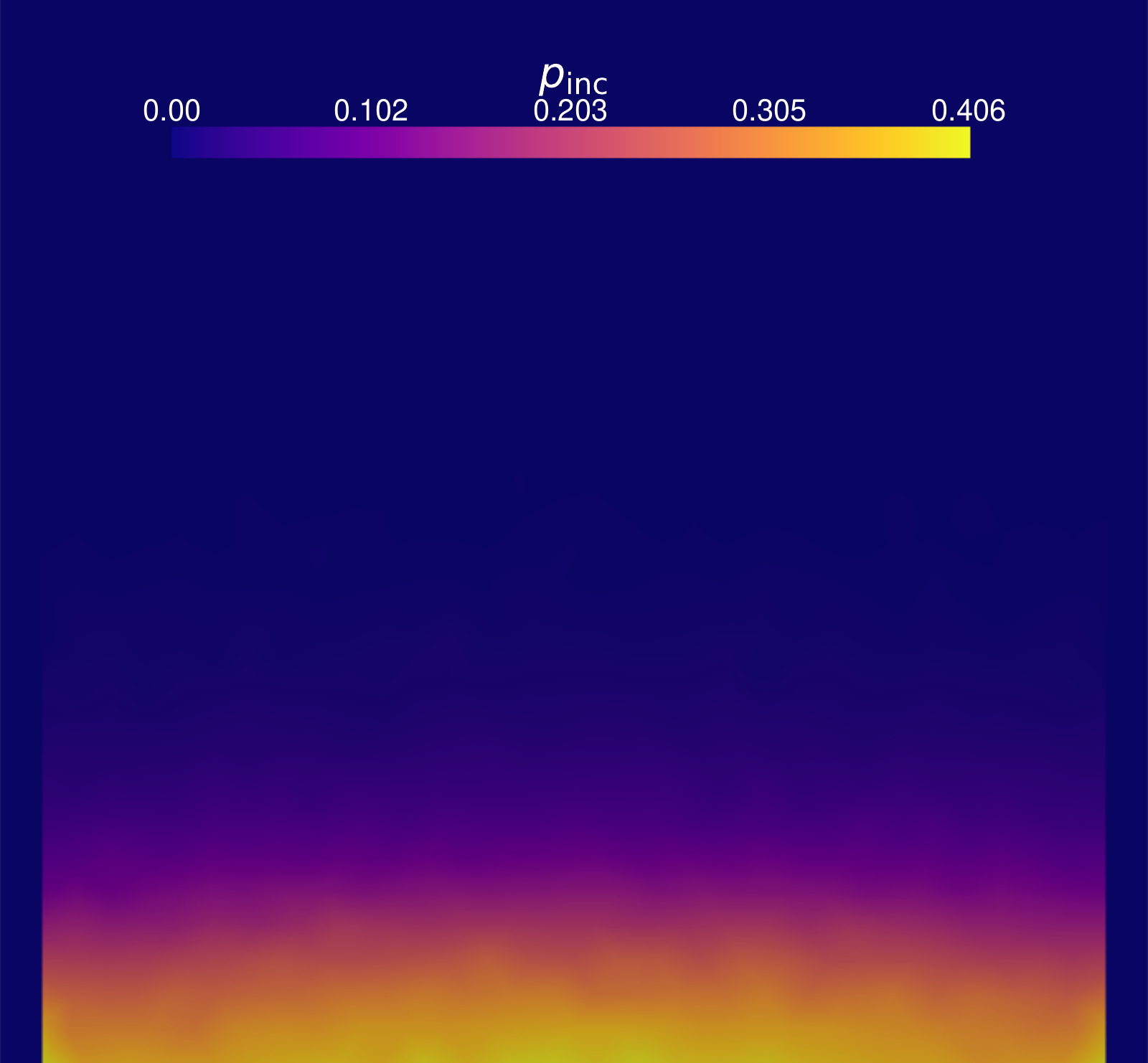}\\[0.2em]
  \includegraphics[width=0.8\linewidth]{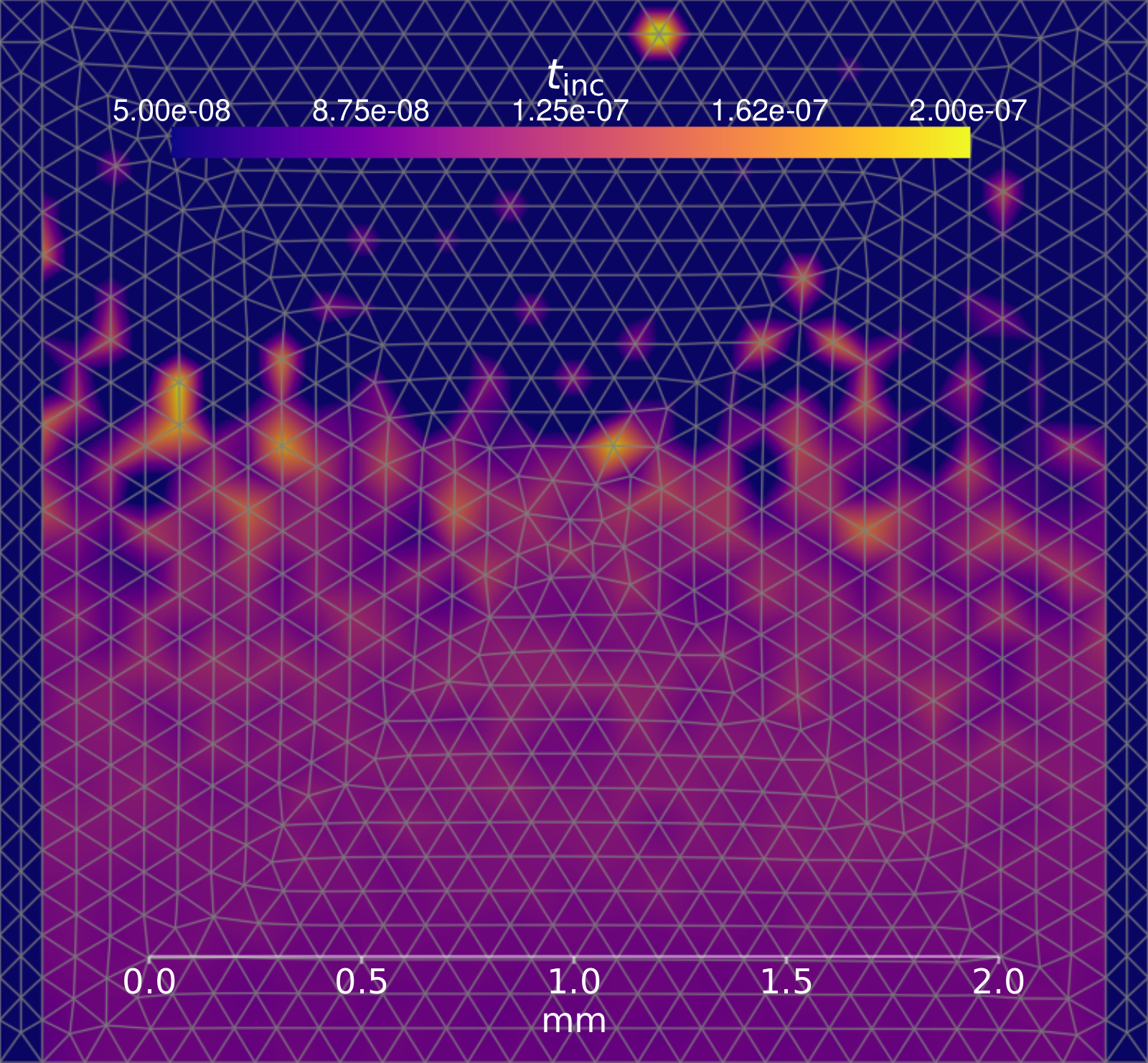}
  \caption{Inception probability $p_\mathrm{inc}$ (top) and inception time $t_\mathrm{inc}$ (bottom) in air with photoionization, using $E_0 = V_0/d = 41.5 \, \mathrm{kV/cm}$. The 2D Cartesian domain has a plate-plate geometry with a separation $d = 2.5 \, \mathrm{mm}$. The mesh is illustrated in the bottom plot. To decrease the noise, the number of runs per initial location was set to $N_\mathrm{runs} = 4000$.}
  \label{fig:pdiv-air-avalanches-pinc}
\end{figure}

Figure~\ref{fig:pdiv-air-avalanches-pinc} shows the inception probability $p_\mathrm{inc}(\mathbf{r})$ in the domain for $E_0 = 41.5 \, \mathrm{kV/cm}$ together with the estimated inception time $t_\mathrm{inc}$.
Since electrons drift up, the highest inception probability is found close to the bottom plate.
The estimated inception times $t_\mathrm{inc}$ are a few tens of nanoseconds for electrons starting close to the bottom plate.
For positions where inception is unlikely, $t_\mathrm{inc}$ is typically larger and also much noisier, since it is then determined from fewer samples.

If the ion secondary electron emission coefficient $\gamma_\mathrm{ion}$ is non-zero at the cathode, the inception voltage can decrease significantly.
For example, with $\gamma_\mathrm{ion} = 10^{-2}$ at the bottom plate we find $E_\mathrm{inc} = V_\mathrm{inc}/d \approx 33.7 \, \mathrm{kV/cm}$, which corresponds to $K^* \approx 4.6$ at the cathode.
If we plug this value into equation~\eqref{eq:gamma-eff} the result is $\gamma_\mathrm{eff} \approx 1.0 \times 10^{-2}$, so that $\gamma_\mathrm{eff} \approx \gamma_\mathrm{ion}$ as expected.

\subsection{Discharge inception around a sphere}
\label{sec:disch-incept-sphere}

We now simulate inception in artificial air around a conducting sphere, with photoionization as a secondary emission process.
As in section~\ref{sec:aval-size-distr}, the sphere has a radius $R = 0.5 \, \textrm{mm}$ and a positive voltage $V_0$.
We use an axisymmetric computational domain, measuring $12 \, \textrm{mm}$ in the $z$-direction and $8 \, \textrm{mm}$ in the radial direction, with the sphere centered at the origin.
On the domain boundaries, the electric potential is set to the free-space solution of the conducting sphere.

The inception voltage corresponding to $\bar{p}_\mathrm{inc} = 10^{-4}$ is $V_\mathrm{inc} \approx 4.73 \, \textrm{kV}$ with our input data.
With this voltage, $K^* \approx 7.4$ at the surface of the sphere.
We remark that the `standard' $K$ given by equation~\eqref{eq:K-integral} is less than $0.1\%$ smaller in this geometry, since most of the electron multiplication occurs in the high field near the sphere.
$K^* \approx 7.4$ results in $\gamma_\mathrm{eff} \approx 6 \times 10^{-4}$ using equation~\eqref{eq:gamma-eff}, which is about 0.2 times the factor $\xi p_q/(p + p_q)$ in equation~\eqref{eq:n-photons}.
This seems reasonable: about half of the photons produced close to the sphere will hit its surface, and \rev{the remaining} photons will, on average, not produce an avalanche of maximal size.

\begin{figure}
  \centering
  \includegraphics[width=0.48\linewidth]{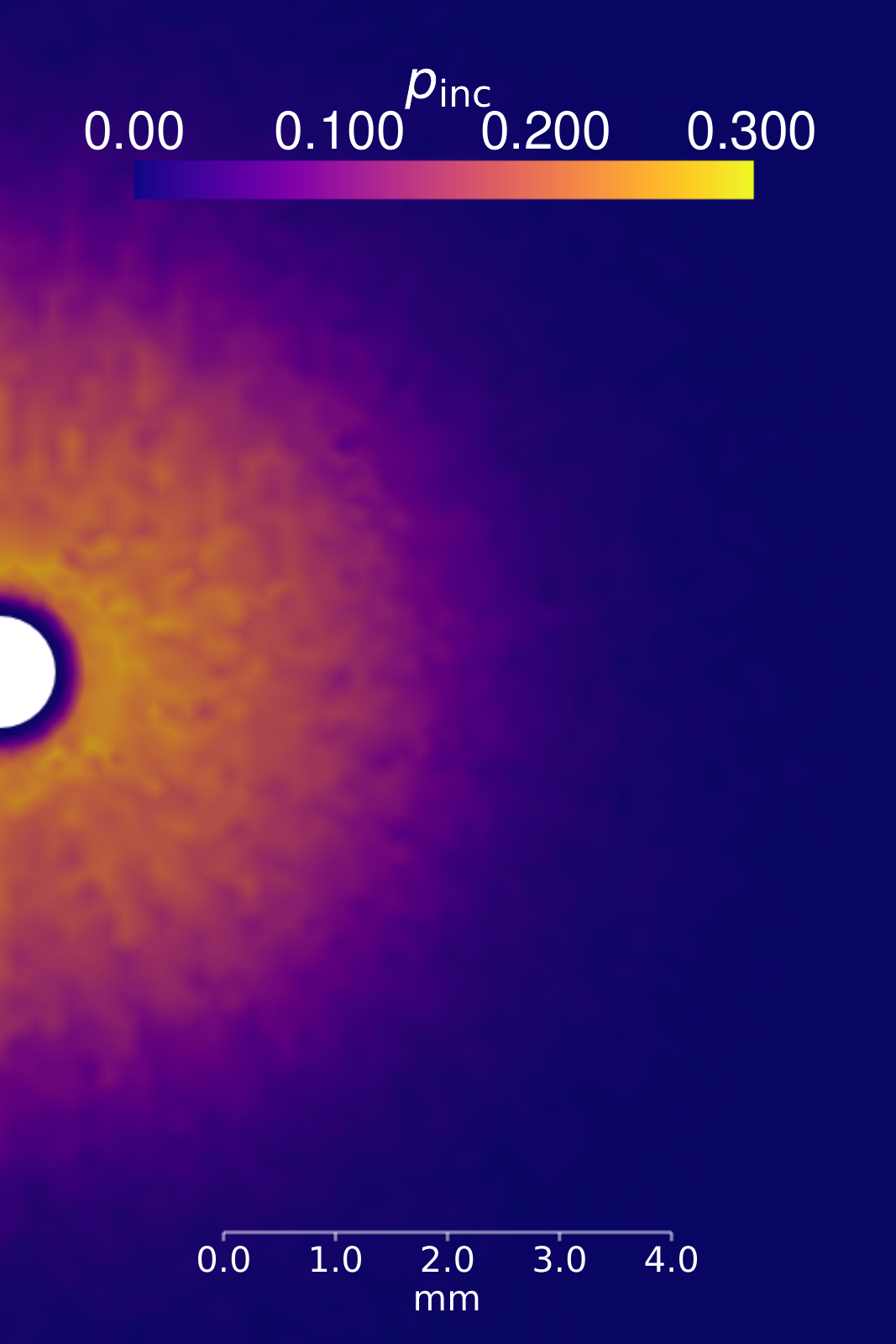}
  \includegraphics[width=0.48\linewidth]{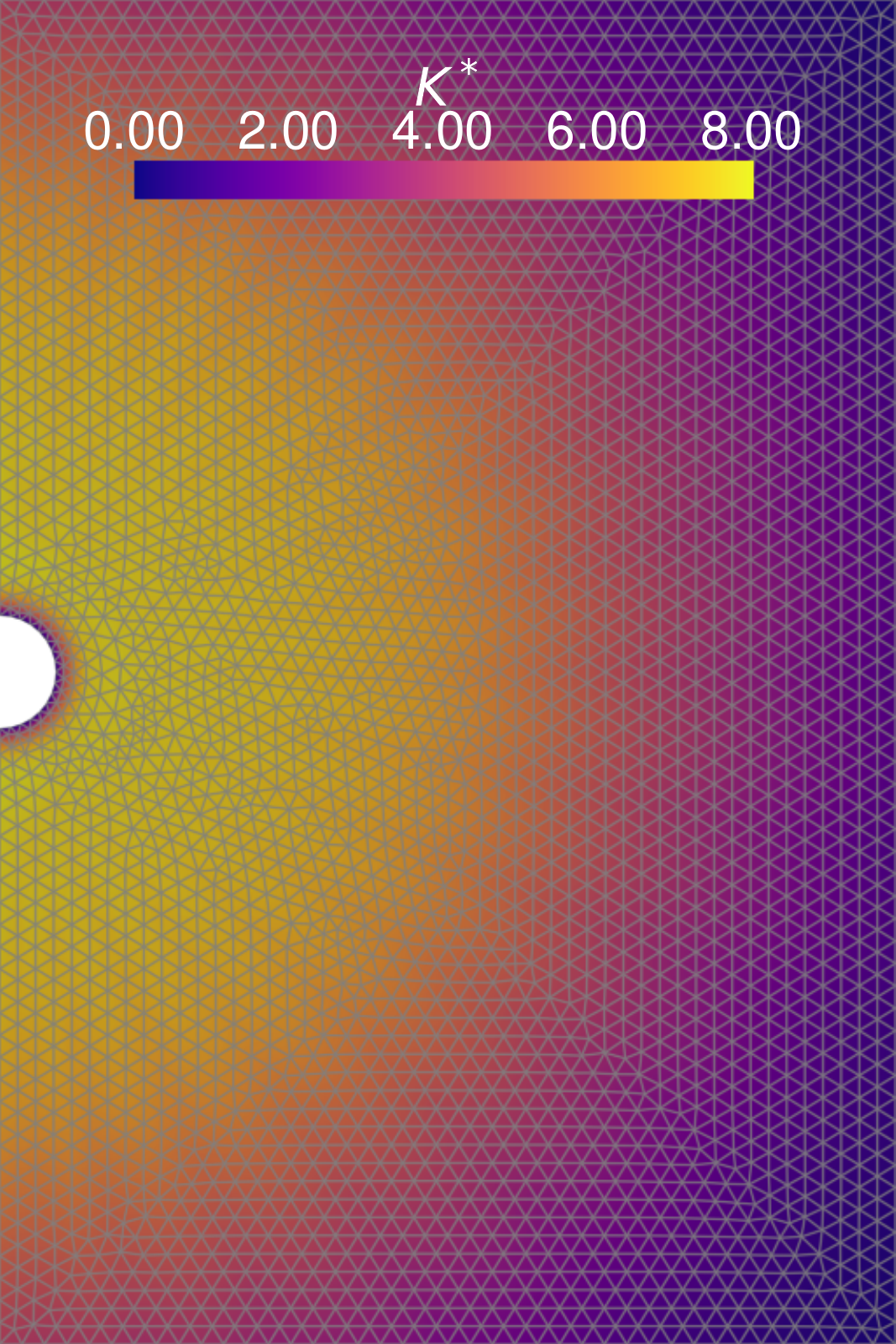}\\[0.2em]
  \includegraphics[width=0.48\linewidth]{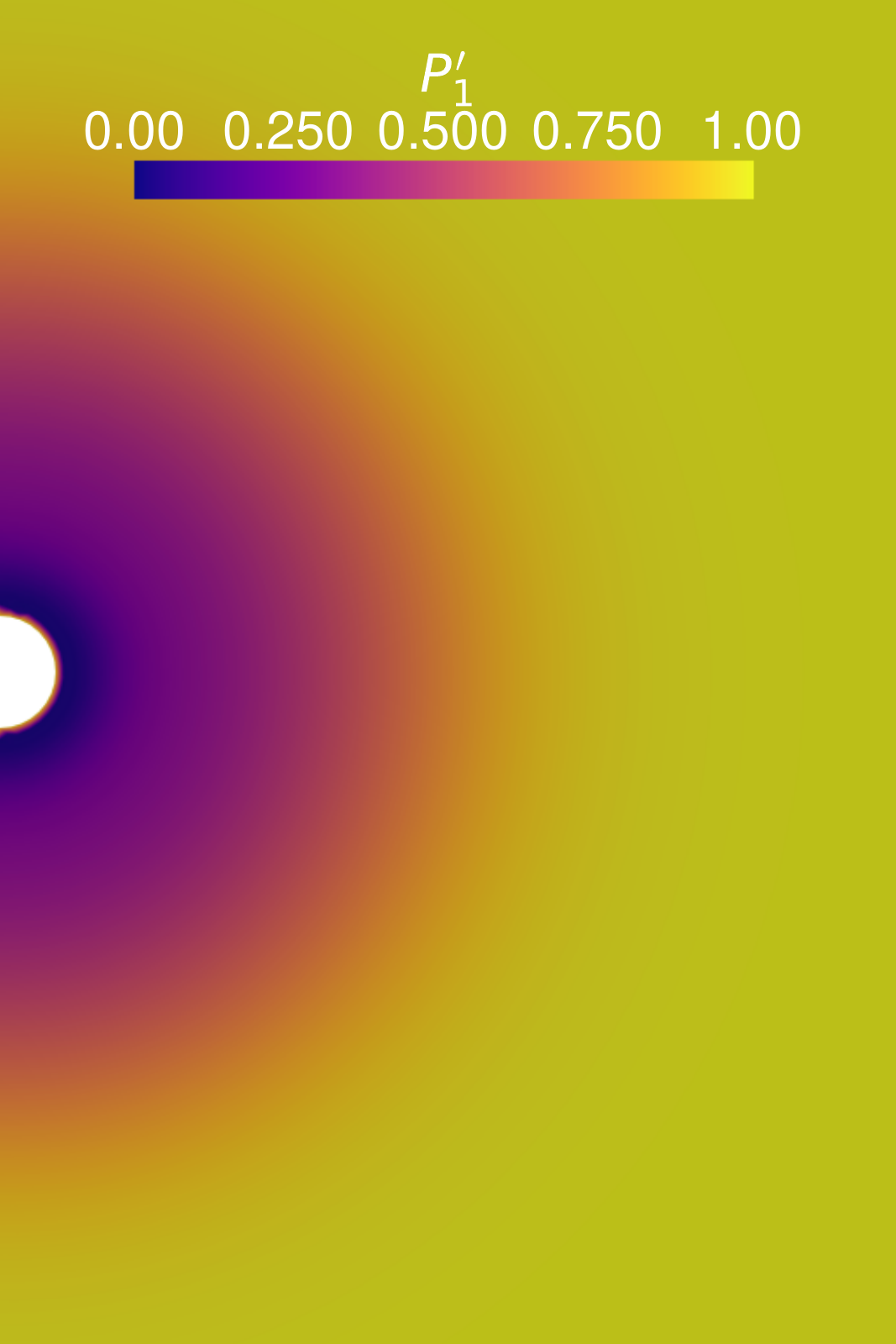}
  \includegraphics[width=0.48\linewidth]{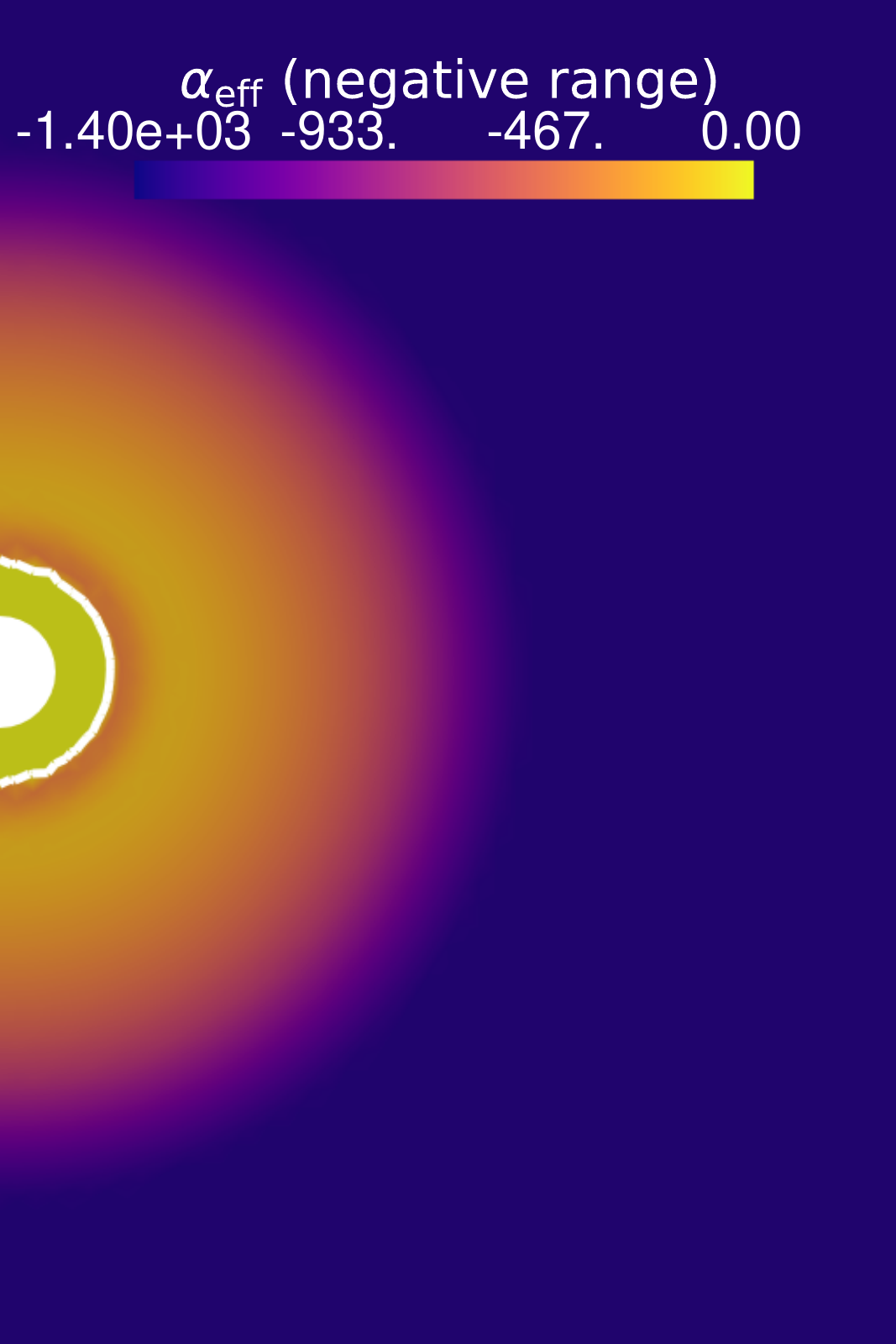}
  \caption{Simulation of inception in air around a spherical electrode at $V_0 = 4.8 \, \mathrm{kV}$, in an axisymmetric domain measuring $6\,\mathrm{mm}$ in the $r$-direction and $12\,\mathrm{mm}$ in the $z$-direction.
    Shown are the inception probability $p_\mathrm{inc}$, the ionization integral $K^*$, the probability $P_1'$ and the effective ionization coefficient $\alpha_\mathrm{eff}$.
    For $\alpha_\mathrm{eff}$ only the negative range is shown; the contour line corresponds to $\alpha_\mathrm{eff} = 0$.
    The number of runs per initial location was $N_\mathrm{runs} = 1000$.}
  \label{fig:pdiv-air-avalanches-sphere}
\end{figure}

Figure~\ref{fig:pdiv-air-avalanches-sphere} shows $p_\mathrm{inc}$, $K^*$, $P_1'$ and $\alpha_\mathrm{eff}$ for $V_0 = 4.8 \, \mathrm{kV}$.
For $\alpha_\mathrm{eff}$ only negative values are shown; the white contour line corresponds to $\alpha_\mathrm{eff} = 0$.
For positions far from the sphere, the probability $P_1'$ indicates how likely it is that the initial electron is lost to attachment.
The figure shows that $p_\mathrm{inc}$ decreases where $P_1'$ increases, and that this coincides with the region where $\alpha_\mathrm{eff}$ becomes strongly negative due to three-body attachment.
However, note that the region from which inception is likely is much larger than the region where $\alpha_\mathrm{eff} \geq 0$.
Close to the sphere $K^*$ decreases since avalanches can only grow over a short distance.

\subsection{Inception with nearby objects}
\label{sec:disch-incept-surfaces}

\begin{figure}
  \centering
  \includegraphics[width=\linewidth]{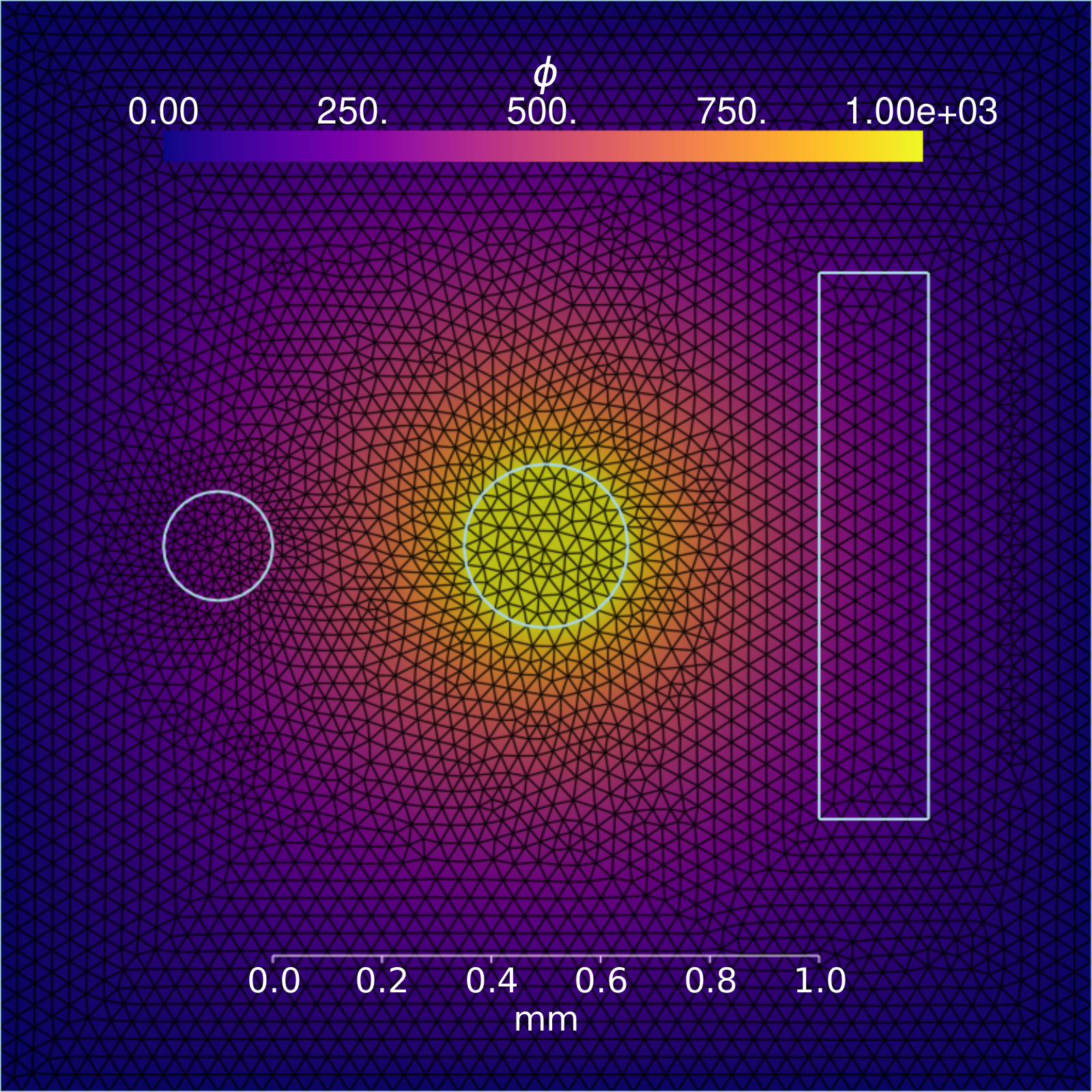}
  \caption{Cartesian 2D domain ($2\,\mathrm{mm} \times 2\,\mathrm{mm}$) with a conducting wire at the center. The electric potential is shown for an applied voltage $V_0 = 1 \, \mathrm{kV}$ at the wire. The smaller circle on the left and the rectangle on the right are dielectric materials with a relative permittivity of one. The domain boundaries are grounded.}
  \label{fig:see-example-domain}
\end{figure}

To illustrate the effect of nearby objects on discharge inception, we have constructed the 2D Cartesian geometry shown in figure~\ref{fig:see-example-domain}.
The central wire is at a positive applied voltage $V_0$.
The dielectric sphere and rectangle on the sides have a \rev{relative} permittivity of one.
They therefore do not affect the electric field, but they do absorb avalanches and photons.
We again use artificial air at 300\,K and 1\,bar, and photoionization is included.

\begin{figure}
  \centering
  \includegraphics[width=0.65\linewidth]{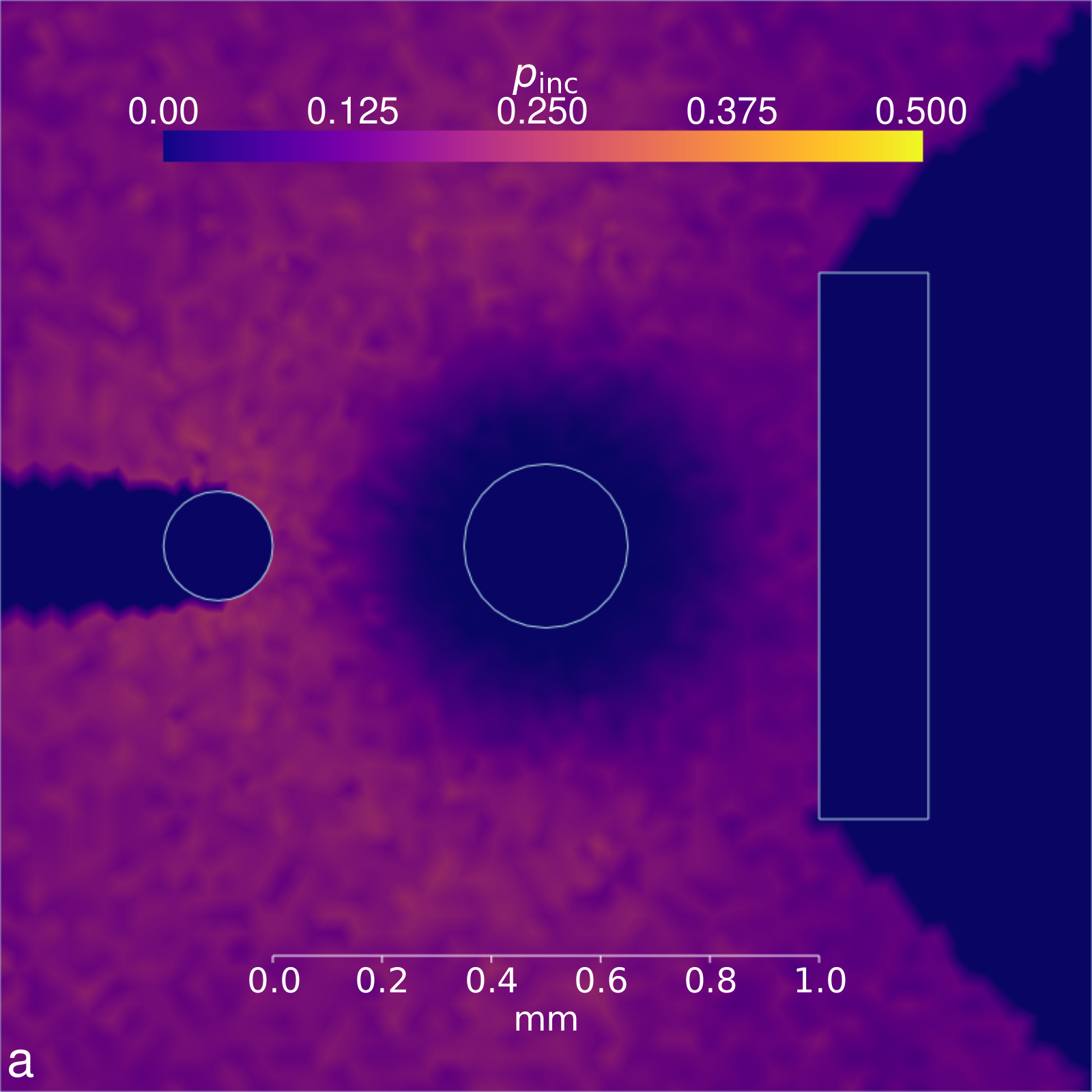}\\[0.2em]
  \includegraphics[width=0.65\linewidth]{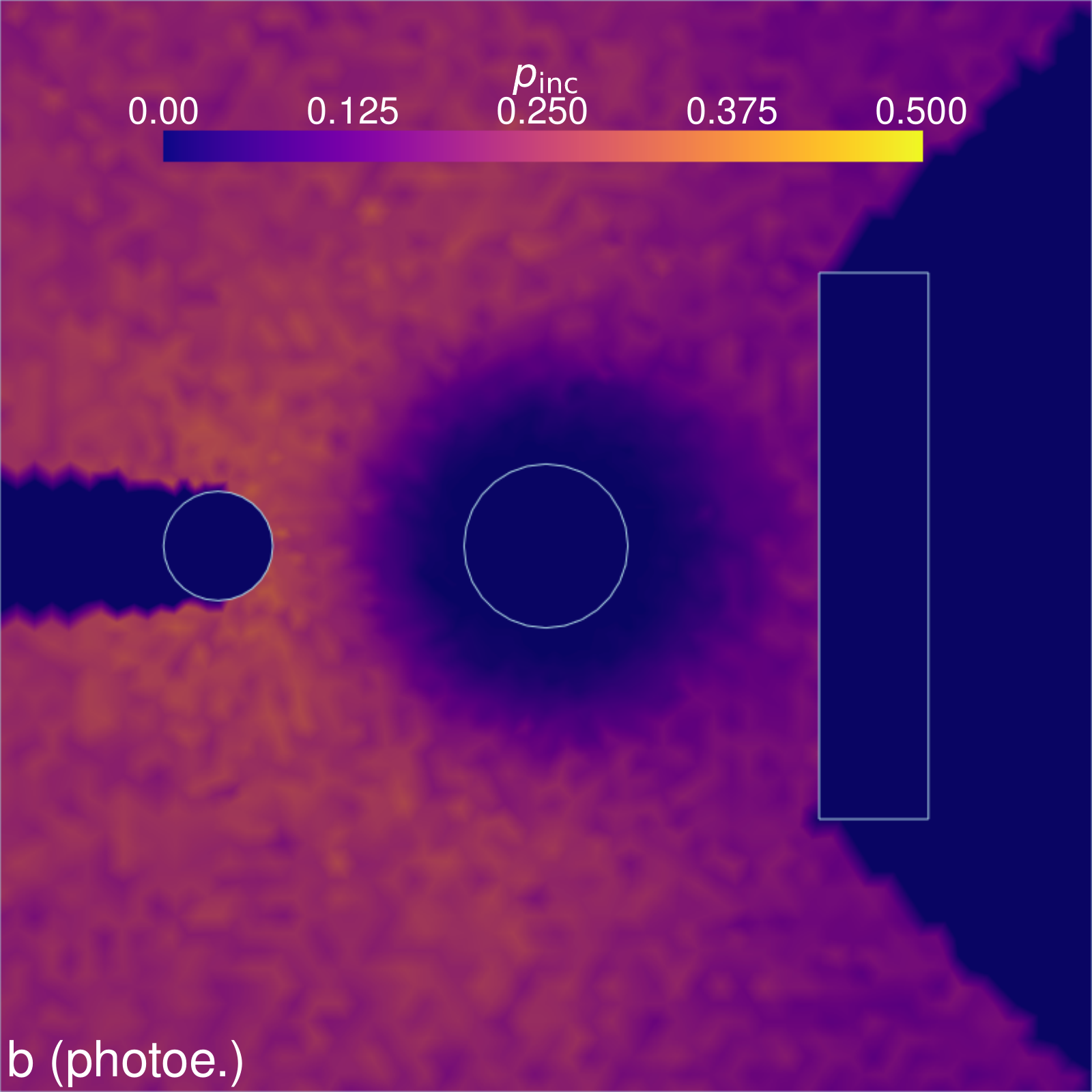}\\[0.2em]
  \includegraphics[width=0.65\linewidth]{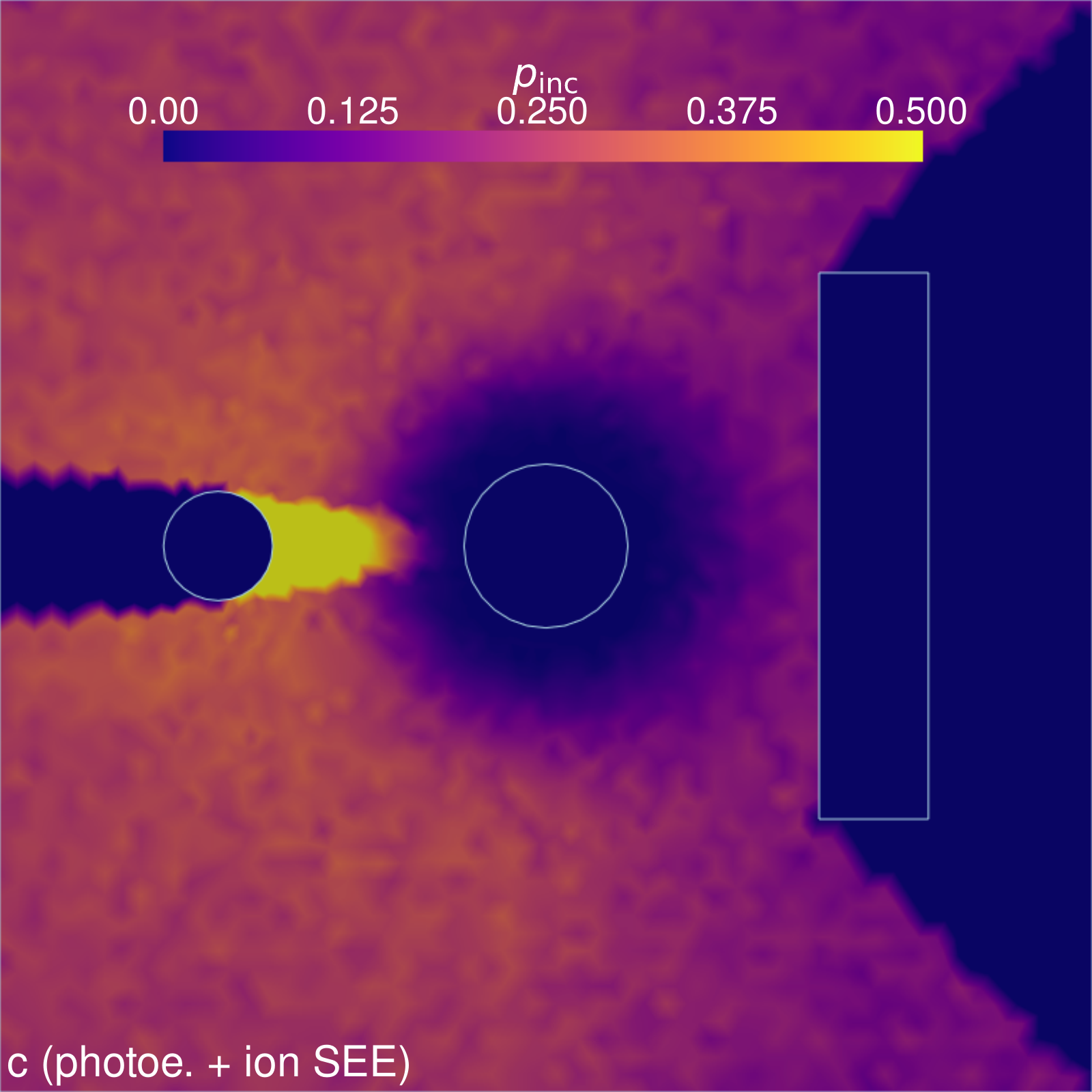}
  \caption{Inception probabilities in air at $V_0 = 3.25\,\mathrm{kV}$, using the domain shown in figure~\ref{fig:see-example-domain}. Three cases are considered: a) only photoionization; b) photoionization with photoelectron emission from the rectangle with $\gamma_\mathrm{surf} = 1.0$; and c) photoionization with $\gamma_\mathrm{surf} = 1.0$ on the rectangle and $\gamma_i = 10^{-3}$ on the left dielectric. The number of runs per initial location was $N_\mathrm{runs} = 400$.}
  \label{fig:see-example-all}
\end{figure}

Figure~\ref{fig:see-example-all} shows the inception probability for $V_0 = 3.25\,\mathrm{kV}$, which is slightly above the inception voltage.
Three cases are considered: a) only photoionization; b) photoionization with photoelectron emission from the rectangle with $\gamma_\mathrm{surf} = 1.0$; and c) photoionization with $\gamma_\mathrm{surf} = 1.0$ on the rectangle and $\gamma_i = 10^{-3}$ on the left dielectric.
In all cases, the inception probability is zero for initial electrons starting behind the dielectric objects.
When $\gamma_\mathrm{surf} = 1.0$ is included, the overall inception probability increases, because fewer photons are `lost' to the rectangular dielectric.
When $\gamma_i = 10^{-3}$ is also included, inception becomes more likely between the central wire and the left dielectric.

Note that only the UV photons that can cause photoionization contribute to photoemission in this example.
Depending on the surface material, lower energy photons might also cause photoemission.
Such photons can be more abundant, and their absorption by the gas will differ from that of the UV photons.
We have not yet implemented a separate photoemission process for such low-energy photons, but for gases without photoionization the photon production and absorption coefficients in section~\ref{sec:photoionization} can simply be changed.

\subsection{Inception in a 3D geometry}
\label{sec:disch-incept-3d}

\begin{figure}
  \centering
  \includegraphics[width=\linewidth]{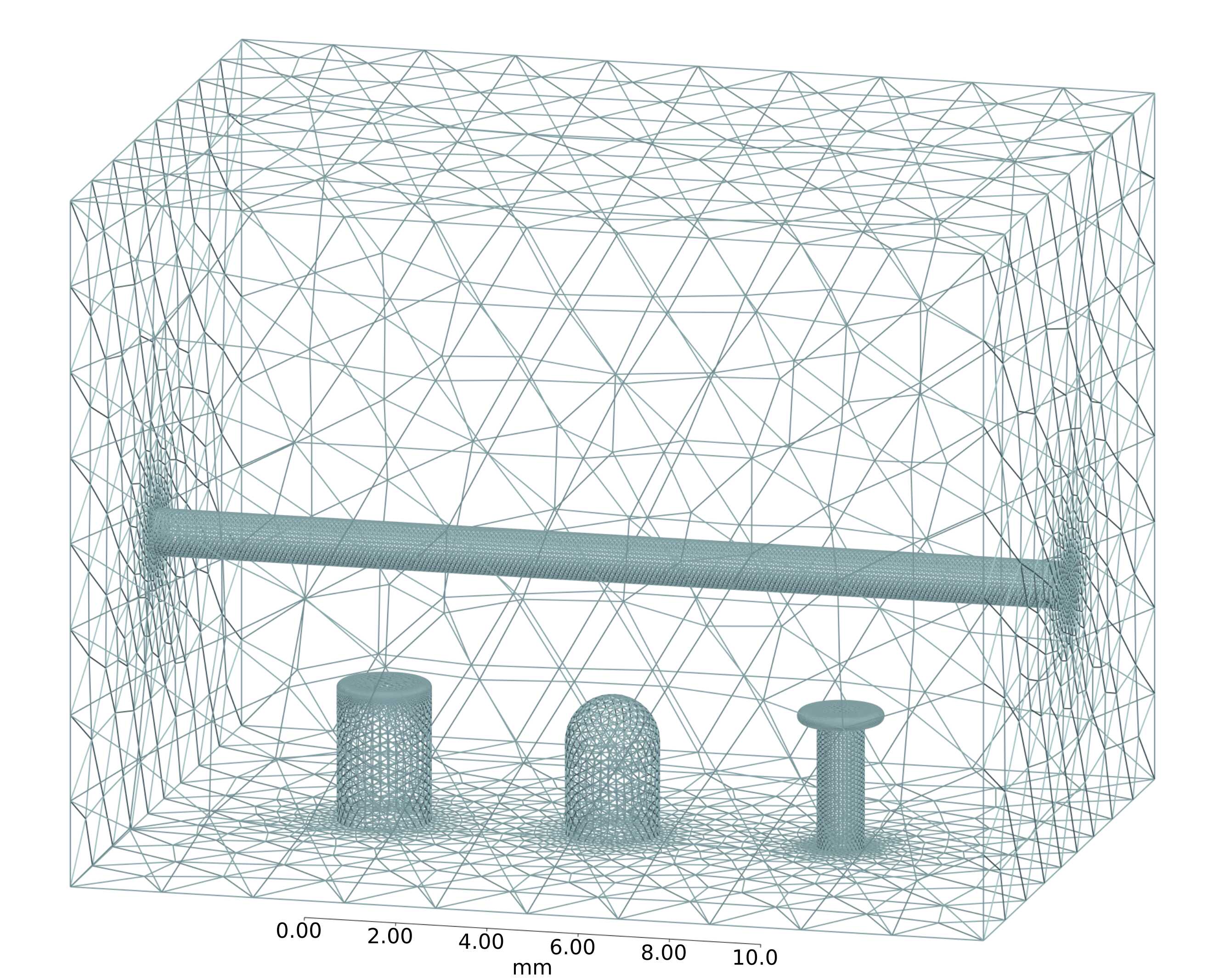}
  \includegraphics[width=\linewidth]{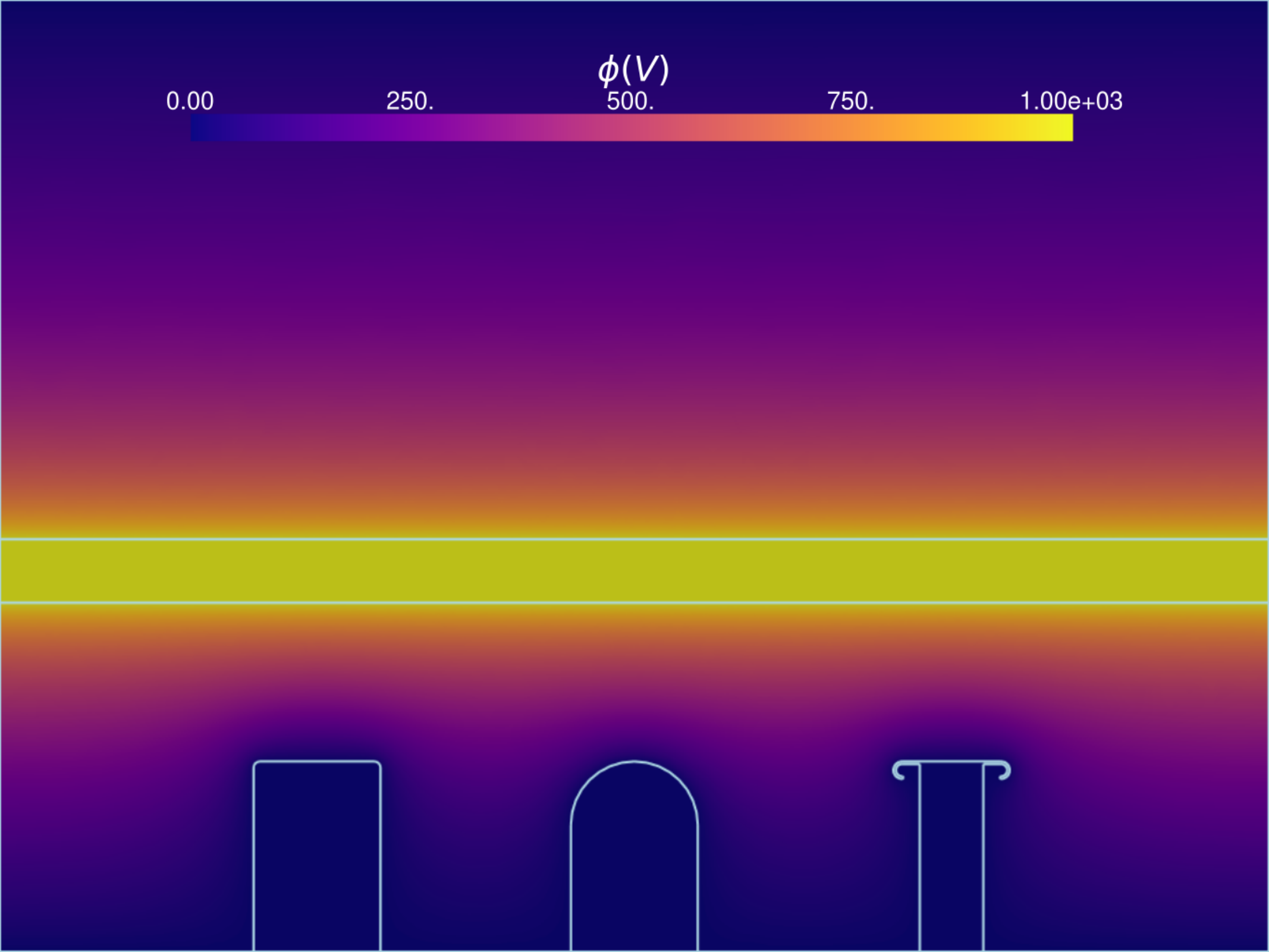}
  \caption{Top: 3D domain measuring $20\,\mathrm{mm} \times 15 \,\mathrm{mm} \times 15 \, \mathrm{mm}$ with a conducting wire. The following boundary conditions are applied for the electric potential $\phi$: Neumann zero on the left and right sides, $\phi = V_0$ on the wire and $\phi = 0$ on all other boundaries. Bottom: a central slice showing $\phi$ for $V_0 = 1 \, \mathrm{kV}$.}
  \label{fig:complex-geom}
\end{figure}

A more complex 3D geometry containing a wire at a voltage $V_0$ is shown in figure~\ref{fig:complex-geom}.
The figure also contains a cross section of the electric potential $\phi$, and the electrostatic boundary conditions are given in the caption.
We again use artificial air at 1 bar and 300 K, and include only photoionization as a secondary electron mechanism.
The inception voltage corresponding to $\bar{p}_\mathrm{inc} = 10^{-6}$ is then about $V_0 = 8.9 \, \mathrm{kV}$ for a positive voltage on the wire, and about $V_0 = -9.2 \, \mathrm{kV}$ for a negative voltage on the wire.

\begin{figure}
  \centering
  \includegraphics[width=\linewidth]{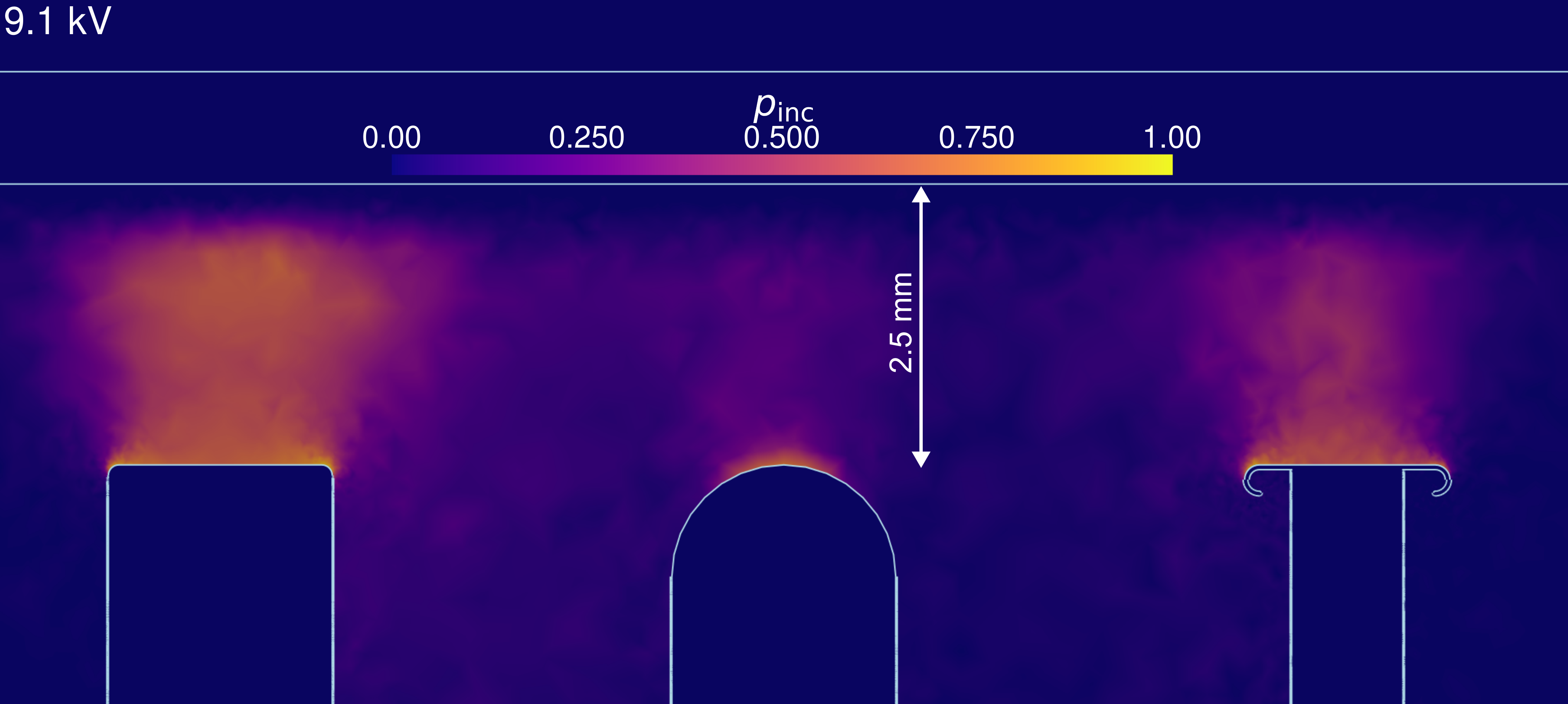}\\[0.2em]
  \includegraphics[width=\linewidth]{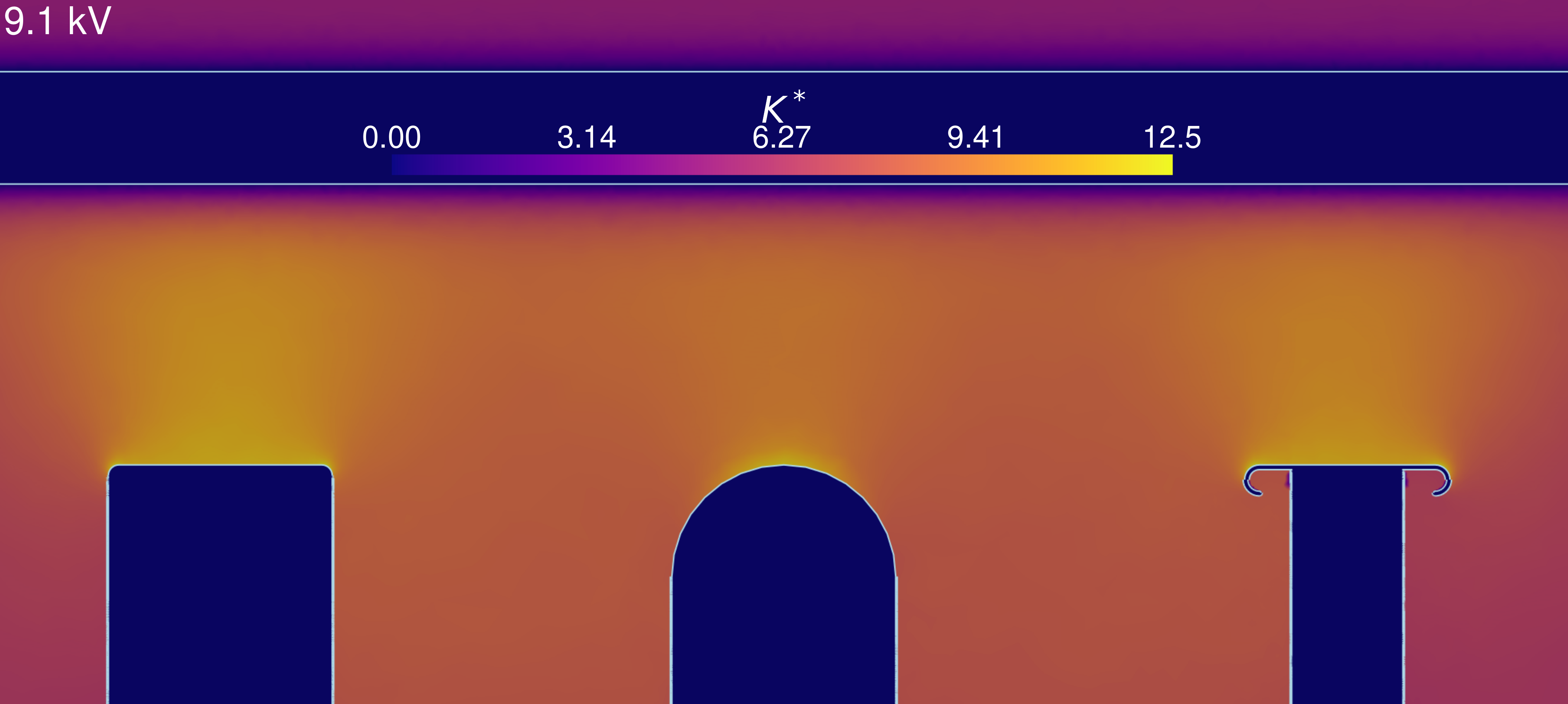}\\[0.2em]
  \includegraphics[width=\linewidth]{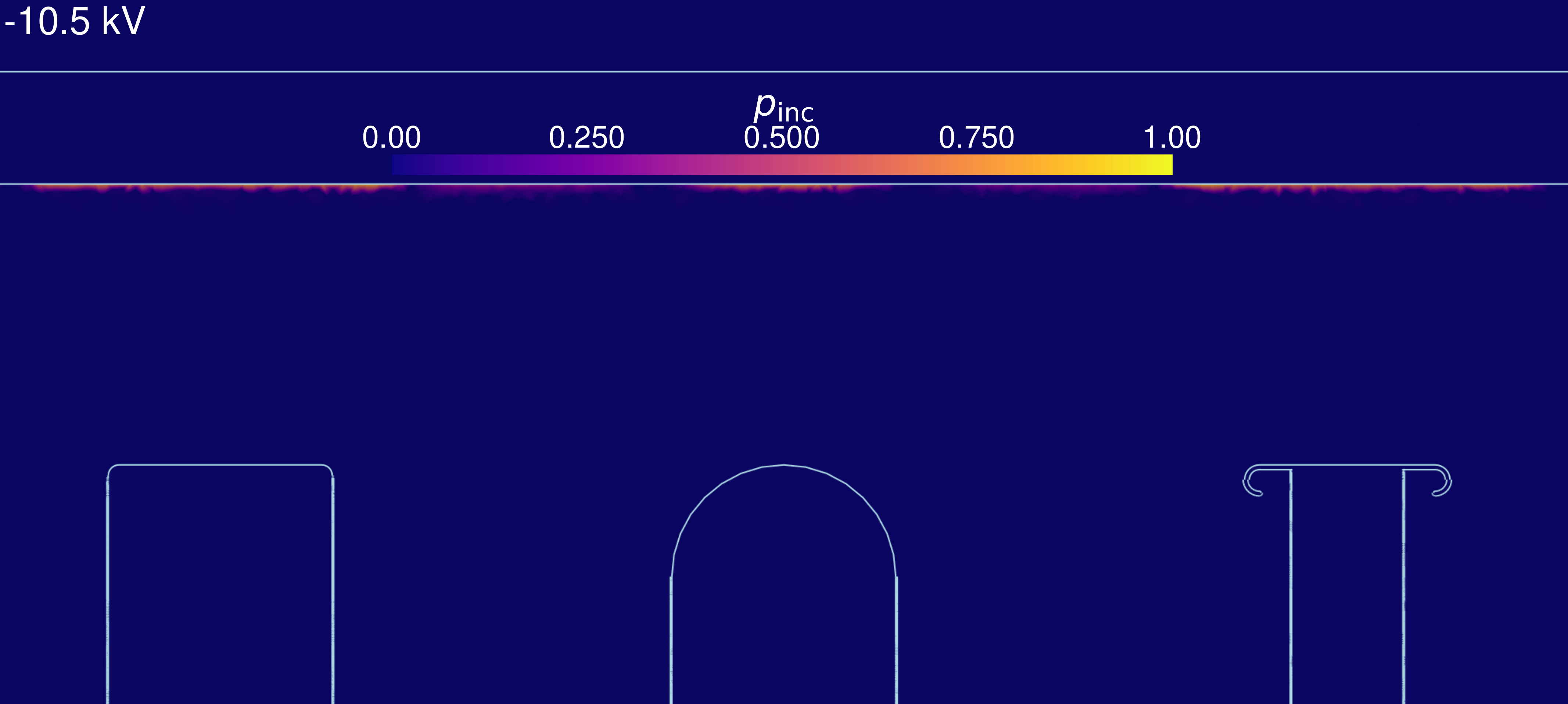}\\[0.2em]
  \includegraphics[width=\linewidth]{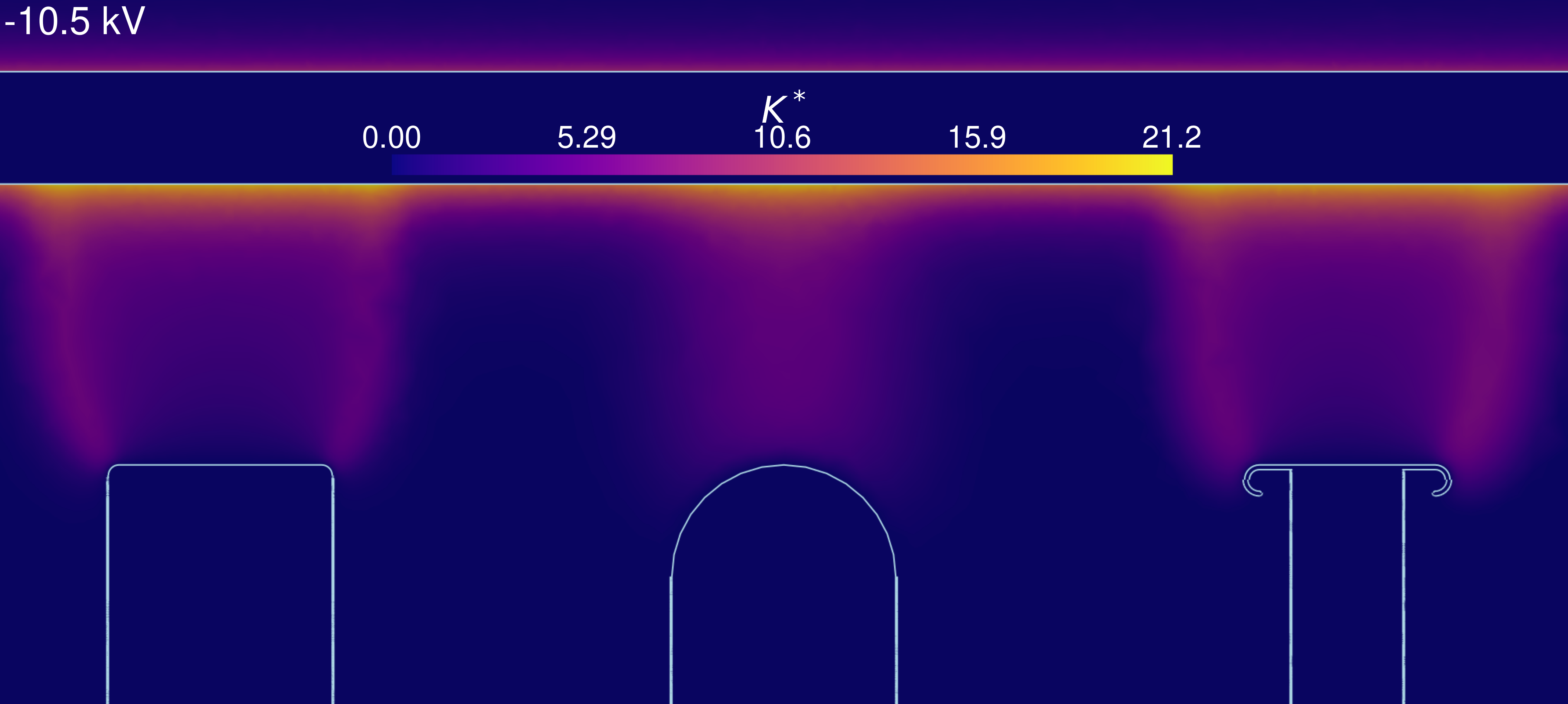}
  \caption{Cross section showing results in 3D geometry with a conducting wire. Top: inception probability $p_\mathrm{inc}$ and $K^*$ for a voltage $V_0 = +9.1 \, \mathrm{kV}$ on the wire. Bottom: results for $V_0 = -10.5 \, \mathrm{kV}$. The number of runs per initial location was $N_\mathrm{runs} = 100$.}
  \label{fig:complex-geom-results}
\end{figure}

Figure~\ref{fig:complex-geom-results} shows $p_\mathrm{inc}$ and $K^*$ for $V_0 = +9.1 \, \mathrm{kV}$ and $V_0 = -10.5 \, \mathrm{kV}$.
For negative $V_0$, a non-zero inception probability occurs in three small regions below the wire, above the protrusions.
This region is the smallest for the central protrusion, and of comparable size for the other two.
For positive $V_0$ inception is likely to happen in a relatively large volume above the protrusions, in particular the leftmost one.
This polarity asymmetry is caused by the fact that the highest electric field occurs around the wire.
The volume where inception is likely is therefore larger when electrons drift towards the wire.
For comparison, the volume-averaged inception probabilities are $4.3\times 10^{-3}$ at $9.1 \, \textrm{kV}$ and $6.0\times 10^{-5}$ at $-10.5 \, \textrm{kV}$.
For a positive voltage inception is more likely to happen above the leftmost protrusion than above the rightmost one.
This is due to the wider base of the leftmost protrusion, which leads to a slightly higher field near the wire.

The noisy pattern in $p_\mathrm{inc}$ seen for negative voltage in figure~\ref{fig:complex-geom-results} is a result of the numerical mesh.
Only electrons starting very close to the wire contribute to the inception probability, since the resulting avalanche size is then likely to exceed the threshold $M_\mathrm{inc} = 10^8$ ($\log(M_\mathrm{inc}) \approx 18.4$) used in equation~\eqref{eq:av-size-threshold}.
In this example secondary electrons are unlikely to be produced very close to the wire because no photoemission process was included.

\subsection{Computational cost}
\label{sec:computational-cost}

For every mesh point several integrals along field lines are computed, see sections~\ref{sec:avalanche-sim-method} and~\ref{sec:ugrid-interp}.
The cost of these integrals depends on the variation in $\alpha$ and $\mu$ along a field line, since this will determine the step size, and it will depend on the number of cells traversed along the field line.
The 3D mesh used in section~\ref{sec:disch-incept-3d} contains about $2.27\times 10^6$ tetrahedra and about $4.35\times 10^5$ points.
With an AMD 2700X 8-core CPU computing all the integrals takes 2 to 5 seconds on this mesh, depending on the applied voltage.
More generally, about $10^5$ points can be handled per second with this CPU for the geometries shown in this paper.

The cost of simulating the avalanches depends on the number of points, the number of runs $N_\mathrm{runs}$ per point, the threshold for the number of future avalanches $n_\mathrm{inc}$ and on the applied voltage.
If at some location avalanches are likely to have a small size, then they are typically not self-sustained and computational costs will be low.
Conversely, if the avalanches multiply over time so that the conditions for inception (see section~\ref{sec:inception-criterion}) are met, the cost will be higher.
The highest cost occurs when the avalanches are just about self-sustained, so that they have to be simulated over a long time before their number either exceeds the inception threshold $n_\mathrm{inc}$ or they all disappear.

The cost of simulating the avalanches is illustrated in table~\ref{tab:comp-cost} for various applied voltages $V_0$, using the 3D mesh from section~\ref{sec:disch-incept-3d}.
A few million avalanches can typically be simulated per second.
With a higher applied voltage there is a larger volume in which self-sustained avalanche growth takes place, which can significantly increase the number of avalanches that need to be simulated.
Note that this effect is much stronger for a positive $V_0$, as can also be seen from figure~\ref{fig:complex-geom-results}.
To speed up computations, it \rev{is} also possible to stop the avalanche simulations as soon as inception has been detected at some location.
With this option enabled, each simulation reported in table~\ref{tab:comp-cost} completed within ten seconds.

\begin{table}
  \centering
  \sisetup{round-mode = figures, round-precision = 2}
  \begin{tabular}{llll}
    $V_0$ (kV) & $\bar{p}_\mathrm{inc}$ & $N_\mathrm{avalanches}$ & Time (s)         \\
    \hline
    $+8.8$     & \num{0.0}              & \num{1.5243462e7}          & \num{0.5138E+01} \\
    $+8.9$     & \num{0.7956E-04}       & \num{4.38290410e8}         & \num{0.2506E+03} \\
    $+9.0$     & \num{0.1202E-02}       & \num{6.28415274e8}         & \num{0.3784E+03} \\
    $+9.1$     & \num{0.4434E-02}       & \num{1.431562759e9}        & \num{0.8772E+03} \\
    \hline
    $-9.5$     & \num{0.0}              & \num{5.715220e6}           & \num{0.2623E+01} \\
    $-10.0$    & \num{0.8001E-06}       & \num{1.2463863e7}          & \num{0.1406E+02} \\
    $-10.5$    & \num{0.6066E-04}       & \num{3.1344934e7}          & \num{0.5258E+02} \\
    $-11.0$    & \num{0.3165E-03}       & \num{6.0400564e7}          & \num{0.1278E+03}
  \end{tabular}
  \caption{Computational cost of simulating inception in the 3D geometry described in section~\ref{sec:disch-incept-3d} for different applied voltages $V_0$ on the wire.
    $N_\mathrm{avalanches}$ is the total number of avalanches simulated and $\bar{p}_\mathrm{inc}$ is the estimate of the volume-averaged inception probability.
    The tests were performed on an AMD 2700X CPU using $N_\mathrm{runs} = 10$, $n_\mathrm{inc} = 1000$ and $M_\mathrm{inc} = 10^8$.}
  \label{tab:comp-cost}
\end{table}

\section{Conclusion and outlook}
\label{sec:conclusion}

We have presented a Monte Carlo method to simulate the inception of (partial) discharges.
The input consists of an unstructured grid with an electric field distribution and electron transport data.
The main outputs are the inception probability and inception time as a function of the initial electron location.
Photoionization is included in the model, and photons can be traced on the unstructured grid to detect when they hit a surface.
Secondary electron emission due to ion impact is also included.
Custom secondary emission coefficients can be defined for different materials in the geometry.

A key component of the model is the rapid simulation of individual electron avalanches.
This was made possible by several developments.
First, we have presented an approximate distribution for the total amount of ionization produced by an avalanche, which can also be used in gases with strong attachment.
Second, we have implemented a fast interpolation library that can be used on unstructured grids.
Furthermore, we assume that avalanches propagate along field lines, which makes it possible to precompute avalanche properties from any starting position in the domain.

We have compared the approximate avalanche model against particle simulations, and generally found good agreement.
Some deviations were observed, which were mostly related to the assumption that ionization and attachment coefficients are a function of the local electric field.
When a secondary electron emission mechanism is included, the model can test whether the number of avalanches grows in time, which is indicative of discharge inception.
We have shown examples of inception simulations in 2D Cartesian, 2D axisymmetric and 3D electrode geometries.

We think that the method presented in this paper can be useful for several applications.
It can be used to interpret inception measurements, in particular the inception probability and time lag, and thereby give information on secondary emission mechanisms and the rate at which free electrons appear.
A tool is provided to determine the inception voltage in a given geometry, but besides this inception voltage, the model will also show in which regions inception is likely.
Compared to earlier work, a distinguishing feature is that photon processes, such as photoionization in air or photoemission from surfaces, can directly be included.

\textbf{Outlook}
As a follow-up to the present paper, we intend to perform an extensive comparison of the new model against experimental measurements.
\rev{Besides inception voltages, we also want to compare inception time lags, which depend on the sources of secondary electrons and on background ionization processes.}

\rev{Several improvements to the model might be required to obtain good agreement with experimental data.
  In dry and humid air it can for example be important to include electron detachment from negative ions, and to model the conversion between different ion species~\cite{verhaartInfluenceWaterVapor1984,pancheshnyiEffectiveIonizationRate2013}.
  In some geometries it could be important to include the effects of electron or ion diffusion, in particular those with sharp electrodes or narrow features.
  Furthermore, for very short gaps, corrections to the local field approximation could be included as well as a mechanism for ion-enhanced field emission~\cite{Farber_2023}.}

\section*{Acknowledgements}

This publication is part of the project REGENERATE with file number KICH1.ST02.21.004 of the research programme NextGen High Tech Equipment which is (partly) financed by the Dutch Research Council (NWO).

\section*{Data availability statement}

The source code of the model described in this paper is available at \url{https://github.com/jannisteunissen/pdsim}

\appendix

\section{Approximating integrals with exponential}
\label{sec:appendix-exp-int}

A number of integrals of the form
\begin{equation}
  \label{eq:exp-integral}
  I(a, b) = \int_a^b \exp[f(x)] \, g(x) \, dx
\end{equation}
have to be computed along field lines, see equations~\eqref{eq:M} and \eqref{eq:P-M1}.
In our implementation, samples of $f(x)$ and $g(x)$ are given along a field line, see section~\ref{sec:ugrid-interp}.
If the electric field is close to homogeneous, the step size between samples can be large.
Since the exponential term can then rapidly vary in a single step, we approximate the integral between samples at $x$ and $x+h$ as
\begin{equation}
  \label{eq:exp-integral-approx}
  I(x, x+h) \approx \tfrac{1}{2} \left[g(x) + g(x+h)\right] \, \exp[f(x)] \, \frac{\exp(k_1 h) - 1}{k_1},
\end{equation}
where $k_1 = [f(x+h) - f(x)]/h$, and where the last term has to be computed in a numerically safe way using an \texttt{expm1} routine.
Equation~\eqref{eq:exp-integral-approx} is based on the assumption that $f(x)$ varies linearly between $x$ and $x+h$, and that $g(x)$ varies slowly.

\printcredits

\bibliographystyle{model1-num-names}

\bibliography{references}



\end{document}